\documentclass[sigconf]{acmart}
\AtBeginDocument{%
  }

\copyrightyear{2025}
\acmYear{2025}
\setcopyright{acmlicensed}\acmConference[CCS '25]{Proceedings of the 2025 ACM SIGSAC Conference on Computer and Communications Security}{October 13--17, 2025}{Taipei, Taiwan}
\acmBooktitle{Proceedings of the 2025 ACM SIGSAC Conference on Computer and Communications Security (CCS '25), October 13--17, 2025, Taipei, Taiwan}
\acmDOI{10.1145/3719027.3765094}
\acmISBN{979-8-4007-1525-9/2025/10}

\usepackage[font={small,it},labelfont=bf]{caption}
\usepackage{graphicx}
\usepackage{xcolor}
\usepackage{booktabs}
\usepackage{xurl}
\usepackage{listings}
\usepackage{multirow}
\usepackage{cleveref}
\usepackage{float}
\usepackage{algorithm}
\usepackage{algpseudocode}
\usepackage{tikz}
\usepackage[subtle]{savetrees}
\usepackage{nth}
\usepackage{balance}

\lstdefinelanguage{EBNF}{
  morekeywords={=,|,{,},[,],;,","},
  sensitive=false,
  morecomment=[l]{//},
}

\begin{document}

\title[What Gets Measured Gets Managed]{What Gets Measured Gets Managed: Mitigating Supply Chain Attacks with a Link Integrity Management System}

\author{Johnny So}
\email{josso@cs.stonybrook.edu}
\affiliation{%
    \institution{Stony Brook University}
    \city{Stony Brook}
    \state{New York}
    \country{USA}
}

\author{Michael Ferdman}
\email{mferdman@cs.stonybrook.edu}
\affiliation{%
    \institution{Stony Brook University}
    \city{Stony Brook}
    \state{New York}
    \country{USA}
}

\author{Nick Nikiforakis}
\email{nick@cs.stonybrook.edu}
\affiliation{%
    \institution{Stony Brook University}
    \city{Stony Brook}
    \state{New York}
    \country{USA}
}

\renewcommand{\shortauthors}{Johnny So, Michael Ferdman, \& Nick Nikiforakis}

\newcommand\ts{\textsuperscript}
\newcommand\todo[1]{\textcolor{red}{#1}}
\newcommand\code[1]{{\small\nolinkurl{#1}}}
\newcommand\codemono[1]{\texttt{#1}}
\newcommand{\todoCite}{\todo{need citation}}

\newcommand\urlArtifact{\url{https://github.com/link-integrity-management-system/lims}}
\newcommand{\parheader}[1]{\smallskip\textbf{#1}}
\newcommand{\circled}[1]{\tikz[baseline=(char.base)]{
            \node[shape=circle,draw,inner sep=1pt] (char) {#1};}}

\newcommand\aliasSystemName{LiMS}
\newcommand\fmtScenarioExpire{\circled{A}}
\newcommand\fmtScenarioChange{\circled{B}}

\newcommand\fmtQueryStatus{\code{queryStatus}}
\newcommand\fmtEvalStageZero{\codemono{No SW}}
\newcommand\fmtEvalStageOne{\codemono{No-op SW}}
\newcommand\fmtEvalStageTwo{\codemono{No-op API}}
\newcommand\fmtEvalStageThree{\codemono{Full}}

\newcommand\paramTrancoTopN{1,000}
\newcommand\paramTrancoListDate{September 18, 2024}
\newcommand\paramTrancoNumPerBucket{200}
\newcommand\paramNumTotalDomains{600}

\newcommand\paramChromeVersion{v128.0.6613.137}
\newcommand\paramNumEvalOverheadTrials{30}

\newcommand\paramCCNumContiguousIndexes{10}
\newcommand\paramCCTotalIndexes{11}
\newcommand\paramCCTimeWindowWeeks{54}
\newcommand\paramCCStartIndex{\code{2023-40}}
\newcommand\paramCCEndIndex{\code{2024-42}}
\newcommand\paramCCFutureIndex{\code{2025-13}}

\newcommand\paramCrawlDateStart{April 02, 2025}
\newcommand\paramCrawlDateEnd{April 09, 2025}
\newcommand\paramCrawlWaitSecondsAfterLoad{5}
\newcommand\paramCrawlRetryHoursOnFail{1}

\newcommand\numUnthrottledDownloadMedianMbps{82}
\newcommand\numUnthrottledUploadMedianMbps{44}

\newcommand\numDomainsExcludedByKeyword{5} 
\newcommand\numDomainsNotConnectable{132} 
\newcommand\numDomainsWithExistingSW{13} 
\newcommand\numPrefilteredDomains{450}
\newcommand\numDomainsMissingStage{54}
\newcommand\numRemainingDomains{394}

\newcommand\numOverheadUnthrottledNoopSW{60}
\newcommand\numOverheadWiFiNoopSW{180}
\newcommand\numOverheadFiveGNoopSW{160}

\newcommand\numOverheadUnthrottledMilliseconds{250}
\newcommand\numOverheadWiFiMilliseconds{400}
\newcommand\numOverheadFiveGLowBandMilliseconds{260}

\newcommand\numCCDomains{85}
\newcommand\numCCLifecycleDomains{10}
\newcommand\numCCLifecycleStableDomains{45}
\newcommand\numCCLifecycleUnstableDomains{40}
\newcommand\numCCLowestRankStableDomains{65}
\newcommand\numCCLowestRankUnstableDomains{20}

\begin{abstract}

    The web continues to grow, but dependency-monitoring tools and standards for resource integrity lag behind.
    Currently, there exists no robust method to verify the integrity of web resources, much less in a generalizable yet performant manner, and supply chains remain one of the most targeted parts of the attack surface of web applications.

    In this paper, we present the design of \aliasSystemName{}, a transparent system to bootstrap link integrity guarantees in web browsing sessions with minimal overhead.
    At its core, \aliasSystemName{} uses a set of customizable integrity policies to declare the (un)expected properties of resources, verifies these policies, and enforces them for website visitors.
    We discuss how basic integrity policies can serve as building blocks for a comprehensive set of integrity policies, while providing guarantees that would be sufficient to defend against recent supply chain attacks detailed by security industry reports.
    Finally, we evaluate our open-sourced prototype by simulating deployments on a representative sample of \numPrefilteredDomains{} domains that are diverse in ranking and category.
    We find that our proposal offers the ability to bootstrap marked security improvements with an overall overhead of hundreds of milliseconds on initial page loads, and negligible overhead on reloads, regardless of network speeds.
    In addition, from examining archived data for the sample sites, we find that several of the proposed policy building blocks suit their dependency usage patterns, and would incur minimal administrative overhead.
\end{abstract}

\ccsdesc[500]{Security and privacy~Web application security}

\keywords{Web Resource Integrity; Policies; Browser; Service Worker}


\maketitle

\section{Introduction}

The Internet is an interconnected web formed by the linking of resources.
By providing core standards such as how to specify a Uniform Resource Locator (URL) for the address of a resource, the web enables a diverse array of applications to interface with one another.
These standards build on one another, relying on their foundations to perform their intended functions, as they provide higher-level, and often more easily programmable features.
However, there currently exists no robust standard that provides adequate integrity guarantees for a web that relies on addresses whose contents can change.

Thus, the external parties which provide resources such as scripts and stylesheets comprise a significant part of the attack surface of web applications.
In supply chain attacks, adversaries compromise existing third parties of a site to modify requested resources to deliver malicious payloads to visitors.
Such attacks have recently taken the form of redirectors injected in a popular polyfill library, affecting hundreds of thousands of sites~\cite{akamai2024polyfillattack}; credit card skimmers injected into a chatbot on an e-commerce site, resulting in a \$1.7M USD fine~\cite{riskiq2018ticketmaster,bankinfosecurity2020ticketmaster}; a skimmer that abused residual trust in an expired domain, affecting dozens of e-commerce sites~\cite{jscrambler2022expiredskimmer}; cryptojackers injected into an accessibility library, with government sites among the over 4,000 affected~\cite{helme2023fiveyears}; fake browser updaters served from blockchains~\cite{guardio2023etherhiding}; and keyloggers in place of trust seals~\cite{bleeping2019trustsealkeylogger}.
Ideally, administrators should discover these attacks immediately after they occur, to minimize the impact on their users. 
However, this may be impractical to do in reality with existing solutions, and so they are often detected after a portion of the site userbase has been exploited.

State-of-the-art integrity mechanisms, limited to only Subresource Integrity (SRI)~\cite{w3c2016sri} and Content Security Policy (CSP)~\cite{w3c2015csp1}, can offer some level of protection against supply chain attacks.
SRI is a web standard that includes a new integrity attribute to the HTML script tag that leverages cryptographic hashes to ensure that the received script content is exactly the same as the expected content.
Similarly, CSP provides features that are relevant to integrity, although it was originally designed to prevent cross-site scripting attacks.
At its core, CSP provides a guarantee that resources are loaded from explicitly allowed origins.
Newer versions of CSP have introduced support for strict content integrity of scripts by leveraging cryptographic hashes to check for exact matches~\cite{w3c2016csp2,w3c2024csp3}.
Both of these standards use strict, hash-based content integrity checks.
However, such integrity verification mechanisms are applicable only for a minority of resources which are not expected to change, such as specific versions of JavaScript libraries loaded from CDNs. 
In all other cases (for both JavaScript as well as arbitrary resources), exact byte-for-byte checks are impractical in real-world scenarios~\cite{so2023more}.

\paragraph{Contributions}
In this work, we propose a Link Integrity Management System (\aliasSystemName{}) to primarily combat supply chain attacks by ensuring the integrity of links on web applications, according to configured integrity policies, at the client in near real time.
In particular, we summarize our contributions as follows:
\begin{itemize}
    \item \textbf{Integrity Policies}: the concept of granular \emph{integrity policies} as methods to describe the integrity of a web resource by declaring (un)expected properties.
    \item \textbf{Policy Enforcement}: the application-agnostic design of an integrity policy verification and enforcement system that blocks HTTPS requests from being sent by clients to resources which have violated their corresponding integrity policies, minimizing the potential for data exfiltration.
    \item \textbf{Centralized Link Management}: the ability to discover all types of links on deployed sites, including first-party, third-party, and anchor links with high fidelity, and manage them from a centralized solution. 
    \item \textbf{Policy Building Blocks}: the proposal of basic policies that can be used as building blocks to form a comprehensive set of integrity policies, while providing guarantees that would be sufficient to defend against recent supply chain attacks detailed by industry reports.
    \item \textbf{Evaluation}: the implementation and evaluation of a prototype, finding minimal performance overhead and no loss of functionality to first-party applications, and demonstrating the applicability of several policy building blocks based on archived snapshots of sites.
\end{itemize}

The rest of this paper is organized as follows: we discuss relevant background in \Cref{sec:02_background}, the system design in \Cref{sec:03_methodology}, the policy building blocks in \Cref{sec:03_blocks}, the evaluation of our prototype in \Cref{sec:04_eval}, the related works in \Cref{sec:05_related}, the limitations and planned work in \Cref{sec:06_disc}, and our conclusions in \Cref{sec:07_conc}.
\section{Background}
\label{sec:02_background}

In this section, we discuss web integrity and the threat model.

\subsection{Resource Integrity}
\label{subsec:02_web}
Subresource Integrity (SRI)~\cite{w3c2016sri} and Content Security Policy (CSP)~\cite{w3c2015csp1} are existing web standards that provide integrity guarantees based on strict mechanisms that check for exact matches between expected and received content, by leveraging cryptographic hash functions.
Through these mechanisms, website developers may specify the expected hash digest(s) of JavaScript files on webpages through the \code{integrity} attribute in HTML \code{<script>} tags for SRI, or through the \code{script-src} directive for CSP. 
If these are present, the user agent (e.g., browser) is expected to compare the hash of the actually received content against the predefined hashes, and block the loading of resources whose computed hashes do not match their expected ones.

Exact matching works for static resources that are not expected to change (e.g., a specific version of a library from a CDN), but does not apply to a significant portion of resources that are dynamic.
For such resources, their URL address, content (e.g., through updates or dynamic modifications to whitespace, syntax, block ordering, comments or data), or dependencies (e.g., fourth-party scripts)~\cite{so2023more} may frequently change.
Furthermore, prior work has uncovered that \emph{frequently-changing scripts are no longer the exception, but the norm in the modern web}: when crawling tens of thousands of domains daily, the study found that only 11\% of script URLs are static, and only 3\% have static content~\cite{so2023more}.
The lack of integrity guarantees for these resources is concerning --- there are no security measures that can protect users if the contents of such resources are unexpected.

\subsection{Threat Model}
\label{subsec:02_threats}

Consider the following scenario: a user visits \code{example.com} which pulls in external subresources on its pages, such as fonts, images, cascading style sheets and JavaScript files.
The administrators of \code{example.com} use an appropriate CSP that defines lists of trusted origins for their external resources, and mark static resources with their expected SRI hash digests.
However, they also use resources that are not easily integrated with strict content integrity checks offered by SRI or CSP, such as \code{foo.com/foo.js}.
Although the \code{example.com} administrators include the origin \code{foo.com} in their CSP \code{script-src} directive, there is no expected hash for that script.

\paragraph{Scenario \protect\fmtScenarioExpire{}: Expiration} 
If \code{foo.com} expires and is re-registered, visitors of \code{example.com} could suffer from a supply chain attack when fetching, and executing, malicious JavaScript from \code{foo.com/foo.js} in their browsers.
Ideally, \code{example.com} administrators would discover this, remove the script from their page, and update their CSP, before the \code{foo.com} domain is expired or re-registered, but this process could take an extended amount of time~\cite{so2022domains}.
Similarly, the domains of anchor links (HTML \code{<a>} tags) found on \code{example.com} are also subject to the same issue.
Although the damage may not be as severe as a supply chain attack (e.g., with the \code{foo.js} script), if an anchor links points to a malicious domain, it will still negatively affect the reputation of \code{example.com}.
SRI and CSP cannot be applied to anchor links. 

\paragraph{Scenario \protect\fmtScenarioChange{}: Link Content Change} 
Alternatively, it may be the case that the domain \code{foo.com} did not expire, but the content of \code{foo.com/foo.js} unexpectedly changes.
SRI and CSP are not applicable because the script is dynamic.
Regardless of whether the change occurs because of a malicious compromise, or an unscrupulous update to its data collection practices, the administrators of \code{example.com} may desire to have an automatic mechanism to inform them if \code{foo.com/foo.js} changes in an unexpected manner, and block that resource for their visitors.
As in Scenario \fmtScenarioExpire{}, site admins may also be interested in applying this mechanism to anchor links as well.

\paragraph{Threat Model}
A user interacting with an online, benign web application that properly utilizes SRI and CSP loads external resources that may have been undesirably modified.
The goal of \aliasSystemName{} is to prevent supply chain attacks on its deployed website, assuming a sufficient and robust set of integrity policies, by detecting ``significant'' changes to dependencies in near real time, blocking all website visitors' requests to these resources, and flagging the incident to administrators.
The threshold for the significance of change is determined by the set of integrity policies that are enabled by administrators, and should ideally encapsulate both Scenarios \fmtScenarioExpire{} and \fmtScenarioChange{}.
If a malicious third-party script is able to execute in the browser of a site visitor, that means the configured integrity policies did not flag the (change in that) resource.
\Cref{subsec:02_trust_model} elaborates on the trust model under which \aliasSystemName{} operates, and \Cref{sec:04_security} further discusses security-critical details that are imposed by design and implementation choices.

\subsection{Service Workers}
In our prototype, we implemented the client of \aliasSystemName{} as a service worker (SW)~\cite{w3c2022serviceworkers}, a performant worker that acts as a proxy between the browser and the network.
Although service workers were not explicitly designed for our use case (one of the original design goals was to enable the creation of offline web applications), we found that their capabilities expressly suited our needs: the ability to transparently intercept \emph{all} HTTPS requests (including navigations) that originate from the pages under its purview, and take different actions.
Furthermore, this implementation choice leverages a web standard supported by all major browsers~\cite{mdnserviceworkers}, and removes the need to directly modify browser code -- which are highly-optimized and massive repositories of software -- thereby improving the accessibility of \aliasSystemName{} for site administrators and researchers.
We further discuss the role of the client SW in \Cref{subsec:02_client} and its disadvantages in \Cref{sec:06_disc}.

\graphicspath{ {./figs/03_methodology/} }


\begin{figure}[t]
\centering
\begin{lstlisting}[
    caption={EBNF grammar for the \aliasSystemName{} integrity policy language.},
    label={lst:ebnf-policy}
]
policy = { rule };
rule = action, url_pattern_page, url_pattern_resource, [ "if", condition_name ], ";" ;
action = "allow" | "deny";
url_pattern_page = url_pattern
url_pattern_resource = url_pattern
url_pattern = '"', { url_char | "*" }, '"' ;
url_char = letter | digit | "." | "/" | ":" | "_" | "-" ;
condition_name = policy_building_block | custom_condition ;
custom_condition = letter, { letter | digit | "_" } ;
\end{lstlisting}
\end{figure}

\begin{figure*}[th]
    \centering
    \includegraphics[width=1.9\columnwidth]{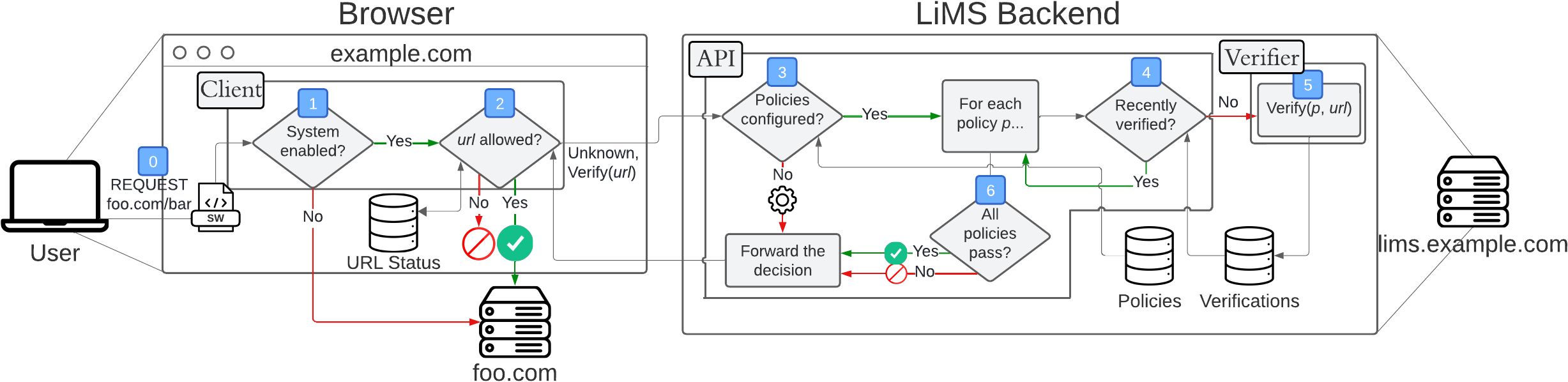}
    \caption{
        Diagram depicting high-level interactions between the client service worker and the \aliasSystemName{} API that comprise the verification protocol to determine whether an outgoing request should be allowed, or blocked, by the service worker.
    }
    \label{fig:diagram_system}
\end{figure*}

\section{Link Management System (\aliasSystemName{})}
\label{sec:03_methodology}

\aliasSystemName{} guarantees that an HTTPS request originating from a user of a website is sent \emph{if and only if} its pre-configured \emph{integrity policies} hold true at the time of, or near the time of, the request.
If at least one policy does not hold true, then \aliasSystemName{} will temporarily block \emph{all visitors' requests} for that resource, and flag the incident for administrators.
This functionality is ensured by three main components working in tandem: the client-side component that intercepts requests (``client''), the server-side component that manages link state (``server''), and the server-side component that verifies configured policies on demand (``verifier'').
\Cref{fig:diagram_system} presents a high-level diagram of \aliasSystemName{} that can be followed with the next sections that discuss:
\begin{itemize}
    \item (\S\ref{subsec:02_policies}) the fundamental integrity policies that describe the (un)expected properties of resources,
    \item (\S\ref{subsec:02_client}) the client service worker that enforces policy decisions and blocks requests to resources that violate their corresponding policies,
    \item (\S\ref{subsec:02_backend}) the server that responds to the client with whether requests should be blocked, as well as the verifier that verifies the configured integrity policies on demand.
\end{itemize}
In addition, the trust model of \aliasSystemName{} is discussed in \Cref{subsec:02_trust_model}, and deployment strategies in \Cref{subsec:03_deployment}.
We open-sourced our prototype implementation of \aliasSystemName{}, which can be found at \urlArtifact{}.

\subsection{Integrity Policy Language Specification}
\label{subsec:02_policies}

Integrity policies form the core of \aliasSystemName{} by expressing certain conditions --- in code --- that are always expected to be true for subsets of resources.
Thus, an individual integrity policy is not intended to encompass all conditions that an administrator wishes to ensure, but rather a single check for a class of resources. 
To this end, \aliasSystemName{} is designed to scale with the use of many concurrently active policies.
Furthermore, as the integrity policies themselves are configurable, the conditions they support are highly varied, running the gamut from exactly matching the content of resources, to verifying the initiator of the request itself, to comparing the network infrastructure of a third party.

\Cref{lst:ebnf-policy} uses Extended Backus-Naur Form (EBNF) to describe the \aliasSystemName{} policy language.
Integrity policies are defined as rule sets that invoke conditions for matching URL patterns.
There are two URL patterns for each policy: the first matching a first-party page URL, and the second matching an expected resource request URL on that page.
The client either allows or blocks each request by matching it against a policy's URL patterns and evaluating the condition (if one is specified).
The condition may refer to the name of a policy building block, as described in \Cref{sec:03_blocks}, or a custom policy condition written by an administrator.
The policy conditions execute on the server and can verify the (un)expected properties of the requested resource, including re-enacting the request context with high fidelity, if desired.
The conditions are evaluated and used asynchronously (\S\ref{subsec:02_backend}).

Administrators can implement policy conditions in a familiar programming language (e.g., NodeJS).
A simple policy can provide the same guarantees as existing defense mechanisms such as SRI (e.g., check that the hash of the response content matches), but the design of \aliasSystemName{} offers greater flexibility.
For example, a condition can replay a script request by launching a headless browser and visiting the first-party page on which the script was found, allowing \aliasSystemName{{} to check parameters such as the script request initiator, the geographic location(s) of the remote provider servers, structural signatures, fourth-party inclusions, and randomization or time-varying behavior~\cite{so2023more}.

However, we recognize that complex security policy systems are not always utilized to their full potential (e.g., CSP is often misunderstood and hard to use correctly~\cite{roth2020complex}), and so accordingly we present universally-applicable policies that can be used as the foundation of comprehensive integrity policies in \Cref{sec:03_blocks}.

\begin{algorithm}[t]
    \small
    \caption{
        Request interception logic from the perspective of the \aliasSystemName{} client-side service worker.
    }
    \label{lst:req_intercept}
    \begin{algorithmic}[1]
        \Procedure{InterceptRequest}{req}

        \State allowed $\gets$ False
        \If{HasValidCacheEntry(req)}
            \State allowed $\gets$ GetStatusFromCache(req) 
        \Else
            \State allowed $\gets$ QueryAndCacheLinkStatus(req)
        \EndIf

        \If{allowed}
            \State Fetch(req) \Comment{Defer caching to the browser}
        \Else
            \State Generate404Response(req)
        \EndIf
        \EndProcedure
    \end{algorithmic}
\end{algorithm}

\subsection{Policy Enforcement}
\label{subsec:02_client}

The \aliasSystemName{} client directly impacts users, as it enforces policies by blocking HTTPS requests to resources that violate their corresponding policies, as shown in the left of \Cref{fig:diagram_system}.
In our prototype, the client is implemented as a service worker (SW), which brings two main benefits to \aliasSystemName{}: transparent operation and performance. 

\paragraph{Transparent Operation} 
When a site deploys \aliasSystemName{}, they will add a JavaScript function call to \code{navigator.serviceWorker.register} that instructs browsers to install the specified SW.
This SW is capable of intercepting \emph{all fetch requests that originate from a page under its control} (including navigation requests); thus, by registering the SW with a control scope set to the root of the domain, it will be able to intercept fetch requests from any page of that domain.
If supported by the browser, the SW will be installed, assume control over all applicable pages, and then refresh all tabs under that domain to ensure that resources on the page only load if their policy verifications succeed.
If not supported by the browser, there will be no change to existing, first-party functionality.

While the client SW is active, it will regularly poll the API server for its configuration settings, doubling in functionality as a heartbeat message.
This periodic polling serves as an alternative for push notifications that the server may send to revoke verification decisions from the client cache (useful if the client disables or is unable to receive push notifications).
If the number of times that the client SW fails to connect to the API server exceeds a configurable threshold, it will automatically revert to a no-op mode by disabling its request interception logic, until it successfully receives a response.

\paragraph{Performance}
In our prototype, we opt for a straightforward implementation that offloads the processing of policies, and detections of violations, to a \emph{server-side component that can reuse the verification decisions for different users}, thereby rendering the client as a simple proxy.
When the client intercepts a request, it only queries the server for whether the request should be allowed to proceed, and locally caches the decision it receives to be reused for a certain duration of time.
An overview of the main client logic can be found in \Cref{lst:req_intercept}.

Server-side verification in this manner enables the following: the first user who loads a resource that needs to be (re-)verified will have the policy verifier make a decision that can be \emph{reused on the next request for that resource, even if it is from a different user}.
Furthermore, this will keep the overhead of the \aliasSystemName{} design within acceptable bounds, even though \aliasSystemName{} intercepts requests --- and sends additional ones --- made by web applications that may have explicit requirements for fast user interfaces.
In short, a server-side component verifies policies and communicates these decisions to client-side, in-browser, integrity proxies that uphold the guarantees for the user.
This choice introduces challenges to accurate verification of integrity policies (which are elaborated in \Cref{sec:06_disc}), but we argue that this is a necessity given modern website-speed requirements~\cite{cloudflarewebsiteperformance} and discuss its implications in \Cref{sec:04_security}.

\begin{algorithm}[t]
    \small
    \caption{
        Policy enforcement logic from the perspective of the \aliasSystemName{} server-side API.
    }
    \label{lst:policy_enforcement}
    \begin{algorithmic}[1]
        \State \textbf{Class} BasePolicy
        \State \ \ \ \ \textbf{Variable} locationSource : String
        \State \ \ \ \ \textbf{Variable} locationTarget : String
        \State \ \ \ \ \textbf{Variable} logic : Function
        \State \ \ \ \ \textbf{Variable} output : Bool
        \\

        \Procedure{GetLinkStatus}{req}

        \State link $\gets$ GetLink(req)
        \State verifications $\gets$ GetVerifications(link.id)
        \For{v of verifications}
            \If{v.failed}
                \State \Return false
            \EndIf
        \EndFor
        \State \Return true
        \EndProcedure
    \end{algorithmic}
\end{algorithm}

\subsection{Link Management and Policy Verification}
\label{subsec:02_backend}

As described in \Cref{subsec:02_client}, the client SW component that enforces policies is intentionally simple --- it defers all decisions it does not know about to the backend.
The backend components, as can be seen on the right side of \Cref{fig:diagram_system}, comprise the API server and the verifier.

\paragraph{\aliasSystemName{} Server}
The server-side component that responds to queries from the SW (``server'') exposes a simple API to determine whether client resource requests should be allowed.
When receiving a request from the client, it records metadata about the request, including the page where the request was encountered and the request URL, determines which policies are applicable to the request, and consults its cache for recent verifications for each policy.
If there exists a valid, recent verification for every applicable policy, the server responds to the client with the decision to allow the request; if at least one policy has recently failed its verification for the request, the server responds with the decision to block the request.
If any policy was not recently verified, the server can respond to the client with a default decision, or optionally schedule the verifier to process it on-demand and use the result, if it does not exceed a configured timeout.
\Cref{lst:policy_enforcement} describes an overview of the main server logic.

\begin{algorithm}[t]
    \small
    \caption{
        Policy verification logic from the perspective of the \aliasSystemName{} server-side verifier.
    }
    \label{lst:policy_verification}
    \begin{algorithmic}[1]
        \Procedure{VerifyLink}{link}
        \State policies $\gets$ GetPolicies(link)
        \For{policy in policies}
            \State success $\gets$ ExecutePolicy(policy, link)
            \State CacheVerification(link, policy, success)
        \EndFor
        \EndProcedure
    \end{algorithmic}
\end{algorithm}

\paragraph{\aliasSystemName{} Verifier}
The last major \aliasSystemName{} component is the verifier, which processes integrity policies to determine if the conditions they express hold true.
These results are termed verification decisions, and they are stored in a cache so that the server can use them to formulate its response to queries from the client.
As shown in \Cref{fig:diagram_system}, the verifier is not directly part of the client-server request flow.
Instead, it interacts with the server asynchronously to verify policies on demand.
If configured, the verifier can also periodically verify all policies before their cached decisions expire, in order to maintain a warm system state and meet performance requirements.

When the verifier receives a verification request, it filters all policies for those that match the page where the request was encountered and the policy target URL pattern, to obtain the set of policies that are applicable to the request.
Next, it executes the logic in each policy and caches the outcome of each verification which remains valid for each policy's time-to-live duration.
These cached verification decisions will then be used by the server, when it receives a request from the client.
\Cref{lst:policy_verification} summarizes the basic verifier functionality.

As the verifier is a server-side component, it is imperative to ensure that the process of policy verification does not sit in the hot path of the client's request to the server.
The asynchronous interaction model between the server and the verifier, where they leverage the database as an intermediary, is designed to minimize the overhead imposed by \aliasSystemName{}.
In an ideal, warm state, verifiers can periodically verify all links against all policies, before any cached decisions expire, to ensure the cache is always populated for requests received by the server.
If there is always a cached decision, the server only requires a database lookup before responding to the client with a decision on whether the request should be allowed.

\subsection{Trust Model}
\label{subsec:02_trust_model}

\aliasSystemName{} is designed to combat the threat model outlined in \Cref{subsec:02_threats}, in which a trusted first-party application depends on resources from a third party whose contents may change without notice, by placing flexible integrity policies on resources. 
To do this, each of the major \aliasSystemName{} components outlined in \Cref{fig:diagram_system} must be trusted: the integrity policies, the client service worker, the server that responds to client queries, the policy verifier, and the database or cache that stores policy information and verification decisions.
If any of these components are compromised or dishonest, then an adversary can effectively disable \aliasSystemName{}.
A dishonest server can respond to client queries to trust undesirable resources; a dishonest verifier can decide to trust resources, regardless of the configured policies; and a compromised database can lead the server to believe that policy verifications succeeded.
Moreover, \aliasSystemName{} necessarily trusts the first-party application (and the client service worker), and expects an honest browser environment.
Otherwise, a compromised first-party may not deliver the expected client service worker code, or a malicious browser extension may silently interfere with regular service worker operations (e.g., by blocking the request that fetches service worker code and preventing installation of the client, or blocking client requests to the \aliasSystemName{} server).

\subsection{Deployment}
\label{subsec:03_deployment}
Interested researchers and administrators can deploy \aliasSystemName{} for existing applications at minimal cost.
Our open-sourced artifact provides ready-made prototypes for each system component as Docker containers.
To self-host a deployment, an administrator needs to obtain a server to host the containers, to expose a public file that contains the service worker code at the root of the sites they wish to protect (e.g., at \code{example.com/sw.js}), and to include a snippet of JavaScript code that registers the service worker in HTML documents.
For sites with existing service workers, the logic of the SW can be added to the start, or to the end, of the request handling logic to avoid breaking existing SW functionality.
As we later see in \Cref{subsec:04_perf}, the immediate benefits for sites that already utilize service workers is actually greater than those for sites that do not, because of the base overhead imposed by the existing service worker request interception.

Service workers are supported by all major browsers~\cite{mdnserviceworkers}, and the enforcement logic is designed to provide smooth opt-in and opt-out experiences.
If a user's browser does not support service workers (e.g., because of an old version), or if the user has configured their browser to block service worker installations (e.g., because of privacy concerns), their browsing sessions will be exactly the same for sites that deploy \aliasSystemName{} and those that do not.
If a user whose browser supports service workers visits an \aliasSystemName{}-enabled site, the SW will force a refresh on all tabs for that site upon installation to ensure that the integrity policies are applied.
If a site owner wishes to remove \aliasSystemName{}, they only need to include a snippet of JavaScript that uninstalls any existing service worker registrations in client browsers, and take down the backend components.
Additionally, if any client SW is unable to reach the backend components, it will automatically revert to a no-op mode that does not enforce any integrity policy decision, nor make additional network requests.

\subsubsection{Multi-Stage Deployment}
Deployment of \aliasSystemName{} can be performed in multiple stages to facilitate a smooth onboarding.
In the first \emph{link discovery} stage, the API server can function in a no-op mode, responding with a default message that allows all requested resources, effectively populating the \aliasSystemName{} database with the links that are requested by users in real time.
In the next \emph{report-only} stage that functions similarly to the mode in CSP, administrators can review the links (as they populate) to start writing their desired policies, and configuring the server to always respond that requests are allowed, even if any policy is violated.
Violations will be reported by the API server and stored in the database by the policy verifier, enabling administrators to review existing policies to check for errors.
When the policies provide sufficient coverage, \aliasSystemName{} can be switched into a normal operation mode: if any corresponding policies are violated, the API server will instruct the client SW to block the request.

\section{Security Considerations}
\label{sec:04_security}

As \aliasSystemName{} is designed to provide additional integrity guarantees, \aliasSystemName{} itself must be sufficiently robust in our threat model.
This section discusses an array of security-critical topics, including the suitability of service workers in our threat model, policy robustness, high-fidelity link discovery and management, caching exploits, camouflaging, and policy consistency and updates.

\subsection{Service Worker}
\label{subsec:04_service_worker}

We leverage service workers to provide integrity guarantees for users, such as those presented in \Cref{sec:03_blocks}, by assuming that the content delivered over the network may be malicious or undesirable because of domain expirations and re-registrations (Scenario~\fmtScenarioExpire{} from \Cref{sec:02_background}) or changes to third-party content (Scenario~\fmtScenarioChange{}).

\aliasSystemName{} uses service workers as the component to enforce policy verification decisions within client browsers.
In our threat model, an undesirably-modified resource in the supply chain of a first-party site should ideally be flagged by one or more policies, causing user requests for that resource to be blocked.
If the resource is not flagged by any policies, then that indicates there is a gap in the set of integrity policies deployed by the site administrators, and the script will be allowed to load.
Any HTTPS requests made by this modified script must pass all appropriate policies, or they will be blocked and flagged.

\aliasSystemName{} requires every page of the first-party site to include the JavaScript code for service worker registration, ensuring that the client browser installs the service worker regardless of which page a user visits.
The service worker only needs to be registered, installed, and activated once, until a new version of the service worker code is fetched.
After installation, the service worker will be present on all subsequent visits.
Cautious administrators would configure their sites to install the service worker before other scripts load (i.e., before redirecting to the main content), as the service worker container is exposed to JavaScript via the \code{navigator.serviceWorker} property.
Before the \aliasSystemName{} service worker is activated, malicious third-party JavaScript could interfere with the service worker registration.
We expect a benign first-use environment and consider this case outside of our threat model, but it is possible for the client to be subject to attacks from a malicious script that was not blocked by the configured integrity policies, which we outline next.

\subsection{JavaScript-based Attacks}
\label{subsec:04_javascript_attacks}

Any JavaScript file that is included and loaded in the webpage will have access to the \code{navigator.serviceWorker} object.
Access to this object grants the ability to install a service worker hosted at an HTTPS URL within the origin; thus, a malicious third-party script cannot install an arbitrary service worker without control of the origin's web server.
A malicious script can uninstall a service worker, but \emph{an uninstalled service worker retains control of the page until the subsequent navigation}.
Thus, malicious third-party JavaScript that was not blocked can attempt to disable \aliasSystemName{} by unregistering service workers and triggering a refresh or navigation, but the \aliasSystemName{} service worker, if properly deployed, will be the first thing re-installed on the subsequent page load.
Theoretically, attackers can repeat this behavior while a gap in the integrity policies persists, but the attack would be limited to denial-of-service (rather than, for example, skimming credit cards), and the behavior would drastically accelerate discovery of the malicious script.

A malicious script that bypasses the configured integrity policies may also attempt a denial-of-service attack on the client service worker by flooding it with spurious messages through the \code{navigator.serviceWorker.controller.postMessage} API, or with fetch requests.
The service worker can easily detect the former.
During normal usage, the service worker would encounter a message only once from the registration script to forcibly refresh the page after installation, and subsequent messages can be ignored.
However, the latter introduces additional network requests and, in turn, additional policy verifications for the resources.

In general, if a malicious third-party script is allowed to load on a client, that means that the configured integrity policies for that site have been bypassed.
\aliasSystemName{} is not intended to defend against other attacks such as prototype pollution or DOM hijacking.

\subsection{Policy and Cache Robustness}
\label{subsec:04_policy_and_cache_robustness}

Sites that deploy \aliasSystemName{} can configure an unlimited number of policies.
As these policies are not communicated outside of the \aliasSystemName{} backend (see \Cref{fig:diagram_system}), a supply-chain attacker cannot directly learn of the policies that are configured for a site.
They can indirectly infer policies by iteratively modifying properties of a resource on the supply chain, and observing whether the \aliasSystemName{} client service worker blocks the request for that resource.
However, \emph{indirectly inferring policies can be expensive}, as many variables are unknown: the number of associated policies for each resource, the exact checks performed by each policy, and the cache duration of each policy.

As such, an adversary that wishes to stealthily modify a recently verified resource, while the decision remains cached, must resort to iteratively probing for information by improperly changing the resource and causing \aliasSystemName{} to block it, for extended periods of time. 
We expect administrators to investigate the resources that \aliasSystemName{} blocks, thereby uncovering the offending link(s) before an adversary is able to extract sufficient knowledge of the policies.
In general, we recommend that administrators configure multiple types of policies to check different integrity dimensions~\cite{so2023more} for dynamic resources in sensitive locations (e.g., the request initiator for, and fourth-party inclusions of, an external script on an ecommerce checkout page).

\subsection{Link Discovery \& Management}
\aliasSystemName{} provides another security-critical feature, in addition to the integrity guarantees: the ability to map out all types of links on their sites with high fidelity, and the ability to manage all of them from one centralized solution.
When the client SW encounters a link with an unknown status, it will query the server to check if the request should be allowed to proceed.
In the process, the \aliasSystemName{} server will automatically build a database of all links, on all pages, of the website.

Typically, link discovery mechanisms can be classified as some combination of the following strategies: crawling, intercepting network traffic (e.g., web application firewall), tracking application or web server log messages, analyzing source code or runtime behavior, or monitoring real users.
There exist many link discovery products in industry, but they are primarily crawlers for the purpose of search engine optimization, web application firewalls, or user behavior monitors for analytics services.
To the best of our knowledge, \aliasSystemName{} is the first proposal to bootstrap integrity guarantees to web browsing through integrity policies, to offer high-fidelity link discovery based on real-user monitoring, and to provide centralized management.

\subsection{Resource Camouflaging}
\label{subsec:04_camouflaging}

The \aliasSystemName{} client does not directly detect camouflaging, as service workers are prohibited from inspecting cross-origin response contents. 
If desired, developers can deploy their own cloaking-detection pipeline that simulates different user classes (e.g., desktop or mobile) and networks (e.g., residential or cloud) to detect links whose contents vary by predetermined parameters. 
This approach was shown effective in prior work~\cite{invernizzi2016cloak}. 
Additionally, developers can also use malware detection services to detect cloaking.  
Such data can feed into a \aliasSystemName{} policy that denies requests to resources that appear to use camouflaging techniques, as in \Cref{app:blocks}, \Cref{lst:policy_threat_intel}.

\subsection{Policy Consistency}
\label{subsec:04_policy_consistency}

In the event that a \aliasSystemName{} administrator erroneously configures contradictory policies, this would cause continuously-failing verifications, and the associated resources will always be blocked.
The administrator would readily notice and investigate the growing number of verification failures. 
It is also possible to incorporate logic to identify conflicting policies and warn the administrator when such policies are enabled.
For example, another monitoring component can be introduced to the backend of \aliasSystemName{}, which can periodically scan for resources that are governed by multiple policies which always result in contradicting verification decisions.

\subsection{Policy Updates}
\label{subsec:04_policy_updates}

If an administrator wishes to update an existing policy, \aliasSystemName{} can invalidate active verification decisions for that policy in the server-side cache.
When the server encounters a client request for a resource governed by the updated policy, it will trigger a re-verification according to the new policy.
Verification decisions cached by the client for these resources will be invalidated either via push notifications or heartbeat responses.

Policies may need to be updated in some scenarios.
For example, if a policy blocks a changed benign resource, the duration of time from when the resource changed until the policy is updated could result in broken page functionality, but administrators can readily investigate blocked resources.
On the other hand, if a policy continues to allow recently-changed resources, a malicious script may be allowed to load, possibly leading to some attacks described in \Cref{subsec:04_javascript_attacks}.
\section{Policy Building Blocks}
\label{sec:03_blocks}

\begin{table*}[t]
    \centering
    \caption{
        Basic integrity policies to use as building blocks to build a comprehensive set of integrity policies.
    }
    \resizebox{1.7\columnwidth}{!}{
        \begin{tabular}{lll}
            \toprule

            \textbf{Policy}                        &
            \textbf{Description}                   &
            \textbf{Recent Incidents}
            \\
            \midrule
            Domain Lifecycle                       &
            Domain was recently registered         &
            \cite{mirrorthief2019ecommerce, jscrambler2022expiredskimmer, malwarebytes2020faviconskimmer, malwarebytes2020faviconexifskimmer,so2022domains}
            \\
            Domain Ranking                         &
            Ranking of domain is below a threshold &
            \\
            Threat Intelligence                  &
            Domain or IP address found in threat
            intelligence feeds                     &
            \cite{guardio2023etherhiding,malwarebytes2020faviconskimmer}
            \\
            Dependencies                           &
            Change in set of third-party
            origins that are contacted by a script &
            \cite{helme2023fiveyears,trendmicro2019adverline,guardio2023etherhiding,mirrorthief2019ecommerce,bleeping2019trustsealkeylogger,sucuri2022wpjstoads,malwarebytes2020faviconskimmer,malwarebytes2019skimmersockets,paloalto2022videoskimmer,confiant2021tagbarnakle,akamai2020skimmersockets,blackberry2023silentskimmer,sansec2020threeyearskimmer}
            \\
            SRI Violation Reporting                         &
            Client-side errors such as failed SRI verifications                                &
            -
            \\
            Infrastructure Attributes                          &
            Geographic restrictions on servers that provide dependencies         &
            \cite{guardio2023etherhiding,blackberry2023silentskimmer}
            \\
            CMS Core File Integrity                             &
            Modifications of core CMS files
            (e.g., WordPress or Magento)
                                                   &
            \cite{sucuri2021ccfile,sucuri2022wpjstoads,sucuri2025corewptheme}
            \\
            \bottomrule
        \end{tabular}
    }
    \label{tab:policy_blocks}
\end{table*}

In this section, we focus on the capabilities of integrity policies, and present several universally-applicable policies in the context of notable historical security incidents arising from malicious changes in resources used by an online, first-party application.
In particular, we focus on how \aliasSystemName{} could have efficiently defended against each breach using simple policies that are not able to be implemented in existing integrity mechanisms (e.g., CSP).

\Cref{tab:policy_blocks} summarizes the policies proposed in this section, and lists related security incidents.
These policies are intended to illustrate their usefulness in the context of actual security incidents: they are not meant to provide an upper or lower bound on the capabilities of integrity policies.
We also note that although the functionality offered by \aliasSystemName{} is a superset of the functionality offered by CSP and SRI, \aliasSystemName{} is not designed to be a drop-in replacement for them.

This section discusses these policies at a high level as the goal is to convey the \emph{potential} of policies, and of \aliasSystemName{}, to provide integrity guarantees in a flexible manner, and not to present exact algorithms.
As such, we refer interested readers to \Cref{app:blocks} for pseudocode representations and to the open-sourced artifact for prototypes of the policies in this section.

\begin{algorithm}[t]
    \small
    \caption{
        A sample integrity policy that denies requests to resources whose domains were recently registered.
    }
    \label{lst:policy_recent_reg}
    \begin{algorithmic}[1]
        \State \textbf{Class} PolicyDomainLifecycle
        \State \ \ \ \ \textbf{Variable} locationSource = "example.com/.*"
        \State \ \ \ \ \textbf{Variable} locationTarget = ".*"
        \State \ \ \ \ \textbf{Variable} logic = IsRecentlyRegistered
        \State \ \ \ \ \textbf{Variable} output = False

        \\
        \Procedure{IsRecentlyRegistered}{req}

        \State threshold $\gets$ GetRegistrationThreshold()
        \State allowlisted $\gets$ IsAllowlistedForRegistration(req.domain)
        \State recentReg $\gets$ GetRecentRegistration(req.domain)
        \State \Return NOT allowlisted AND recentReg $>$ threshold
        \EndProcedure
    \end{algorithmic}
\end{algorithm}

\subsection{Policy: Domain Lifecycle}
\label{subsec:04_policy_domain_lifecycle}
It is imperative for administrators to be able to detect when their included resources belong to domains that are about to expire, or have already expired, as their sites may be vulnerable to supply chain attacks by malicious re-registrants of the expired domains~\cite{so2022domains}.
However, there are no default or standardized mechanisms that perform this function.
It is possible for dependencies of domains to expire and go unnoticed for months, and for attackers to select particular expired domains to target infrastructure, or opportunistically re-register them and exploit whatever is available~\cite{so2022domains,linksentry_expired}.
For instance, a 2022 report detailed a campaign that re-registered the expired domain of an analytics service that was discontinued in 2014 to serve credit card skimmers, and it was still able to impact over 40 different e-commerce sites~\cite{jscrambler2022expiredskimmer}.
There have been many other campaigns where attackers infect websites, inject malicious code on checkout pages, and exfiltrate credit card numbers as users type them, to newly-registered domain names~\cite{mirrorthief2019ecommerce, jscrambler2022expiredskimmer, malwarebytes2020faviconskimmer, malwarebytes2020faviconexifskimmer}.

\Cref{lst:policy_recent_reg} presents a basic policy that blocks requests for all resources whose domains were registered after a configured threshold.
Recent registration data can be obtained from WHOIS data, or inferred from passive DNS data.
Although the policy itself prevents requests to domains that are newly registered and does not attempt to distinguish domains that had previously expired, it can be modified to distinguish between previously-observed domains that have expired, and then were re-registered.
This policy does not protect against attackers who have compromised an existing party in the supply chain of the first-party site, but it does effectively negate the threat of re-registrations of existing domains in the supply chain that were left to expire (Scenario~\fmtScenarioExpire{} from \Cref{sec:02_background}) and the common DNS evasion pattern that uses throwaway, short-lived domains in web malware infrastructure~\cite{antonakakis2010building,bilge2011exposure}.
Note that this policy also applies to domains that exist in CSP allowlists.

\subsection{Policy: Domain Ranking}
\label{subsec:04_policy_domain_ranking}
Domain reputation and popularity can serve as an indicator of quality and security: higher-ranking sites are generally expected to have better security postures because of their amount of users.
Although it is not an absolute relationship, studies have found some evidence that generally support this correlation~\cite{roth2020complex,silva2024worldwide}.
Thus, administrators may desire restricting resources that are fetched from, and linked to via anchor links, lower-ranked domains for security, or to improve the ranking of their own domains.
\Cref{app:blocks}~\Cref{lst:policy_reputation} encodes this logic as an integrity policy, denying requests for resources to domains with a ranking that is below a configured threshold, and not explicitly allowed.
This policy does not protect against attackers who can introduce a high-ranking domain, but adversaries commonly resort to using throwaway domains, which will inevitably be low ranked or unranked in robust ranking lists (e.g., Tranco~\cite{pochat2019tranco}).

\subsection{Policy: Threat Intelligence}
\label{subsec:04_policy_threat_intel}
Threat intelligence feeds and domain blocklists provide data sources comprising indicators of compromise.
These sources aggregate suspicious indicators from prior security incidents, such as data exfiltration endpoints (domains or IP addresses), or hashes of compromised files, and are often integrated into security infrastructure.
Similarly, malware detection services (e.g., VirusTotal) check for not only signatures of previously-identified malicious content, but also suspicious behavior through static and dynamic analyses.
Thus, administrators may desire to use such services to scan the external resources they are linking to from their first-party site, while maintaining their own cache of previously-scanned content to minimize API calls.
\Cref{app:blocks}~\Cref{lst:policy_threat_intel} incorporates this idea into one \aliasSystemName{} policy, enabling the verifier to automatically block requests for domains that appear on threat intelligence services, or whose contents were flagged as suspicious by static or dynamic analyses.

One method to minimize costs of external scanning is to check files whose contents have changed, particularly for file types for which SRI is not applicable, such as images and audio files.
The metadata of image files may be used for covert payload delivery or data exfiltration.
Malwarebytes has published a number of reports detailing adversaries masquerading skimmers as favicons: in one 2020 incident, the favicon request URL would serve a web skimmer if the word ``checkout'' was in the Referer header~\cite{malwarebytes2020faviconskimmer}, and in another, the skimmer payload was hidden in the EXIF metadata field of a favicon~\cite{malwarebytes2020faviconexifskimmer}.
In contrast, maliciously-crafted audio files may trigger arbitrary code execution when browsers process them~\cite{xie2021webaudiosafari,cve2020webaudiosafari}.
 
This type of policy can also be designed with heuristics to identify common behavioral patterns in supply chain attacks, by leveraging first-party knowledge and resources.
For example, it is common for obfuscated, malicious snippets of JavaScript to be injected into otherwise-benign code, which is sometimes called a \emph{benign-append} attack in industry reports.
In 2018, a common JavaScript library loaded by 4,000 sites, many of which belonged to governments, was injected with a cryptojacking snippet~\cite{helme2023fiveyears}; in 2019, an advertising agency was compromised and delivered credit card skimmers to 277 e-commerce sites~\cite{trendmicro2019adverline}; in 2023, 510 WordPress sites were found embedded with bridgehead code that retrieved second-stage fake browser update payloads from a malicious smart contract, effectively leveraging the principles of blockchains to serve as bulletproof hosting~\cite{guardio2023etherhiding}.
In all of these incidents, obfuscated, malicious code was injected into existing resources.
Policies that leverage first-party knowledge will know whether certain scripts were previously obfuscated, or expected data values in analytics scripts (e.g., the ID used by the Google Tag Manager loader script that determines what further scripts are pulled into the site).
Another targeted behavioral pattern could be the tendency for attackers to reuse or revive older infrastructure that have been previously marked suspicious; an incident in 2020 involved a credit card skimmer that masqueraded as a favicon, and the IP address of the domain that served the skimmer had been flagged malicious three years prior~\cite{malwarebytes2020faviconskimmer}.

\subsection{Policy: Dependencies}
\label{subsec:04_policy_dependencies}
Scripts may regularly contact different origins (or URLs) during their execution.
\aliasSystemName{} supports the ability to check for changes to the set of \emph{URLs} contacted by individual \emph{resources}, as opposed to the granularity of \emph{origins} contacted by individual \emph{pages} as offered by CSP.
Consider a scenario in which script \code{foo.js} has changed and contacts an origin \code{bar.com} that it never contacted before.
If \code{bar.com} is already contacted by a different script on the page, and thus present on the CSP allowlist, this issue may never be flagged to administrators without \aliasSystemName{}. 
Or, it might be the case that \code{foo.js} contacts a different URL on a domain \code{bar.com} that it has previously contacted.
This would be particularly concerning if \code{bar.com} allows for user-provided scripts to be uploaded.
For example, Google Tag Manager (GTM), a popular analytics library, enables developers to host their own set of custom scripts, differentiated by a single ID parameter in the URL and in a variable in the initial loader script content.
Adversaries are known to abuse brand trust and change the parameter in the GTM script URL that controls which scripts are loaded~\cite{sucuri2018magecartwithgtm,sucuri2023magecartwithgtm,sucuri2025magecartwithgtm}.

\Cref{app:blocks}, \Cref{lst:policy_dependencies} specifies a straightforward version of this policy.
In production, this policy may not be readily applicable for dynamically-generated URLs.
For example, if a third-party script contacts dynamically-generated subdomains of a fourth party, the policy may have to be adjusted to consider only the eTLD+1 instead of the full domain name, if all subdomains are controlled by the same entity.
However, if different subdomains redirect to different user-controlled content, it would be prudent to consider the full domain name.

\subsection{Policy: SRI Violation Reporting}
\label{subsec:04_policy_sri_violations}
One challenge faced by web developers may be the lack of visibility into silent errors encountered by visitors.
For example, although CSP offers a built-in reporting mechanism for resources that were blocked because of violations, SRI does not.
When a script is blocked from loading because of an SRI violation, the browser only reports the failure in its developer console.
The closest feature to a reporting mechanism for SRI is a now-deprecated proposal to include \code{require-sri-for} in CSP, which would block scripts without an SRI \code{integrity} attribute and leverage CSP \code{report-to}~\cite{udncsprequiresrifor}.

Administrators can use a \aliasSystemName{} policy to ensure that scripts that do not match their expected SRI digests will provide violation reports to administrators, in addition to providing the same functionality as SRI if desired.
\Cref{app:blocks}~\Cref{lst:policy_visibility} describes a policy that provides visibility into these errors that are, by default, only visible in the developer console for browsers.
Although it does not directly provide additional integrity guarantees for users, it improves the ability for administrators to detect problems with critical resources, and thus minimize the response time during attacks.
Additional client-side errors that might be of interest to developers are TLS connection errors for subresources on the page --- when the browser encounters such errors, they are also only reported in the developer console.

\subsection{Policy: Infrastructure Attributes}
Depending on the application infrastructure and the domain context, administrators may need to make policies that encode restrictions or business logic, such as where the physical remote servers of dependencies are expected.
For example, a common deployment pattern geographically distributes hosting servers, and directs users to servers that are physically close to their location to minimize latency.
In addition, security-conscious developers may be concerned about dependencies from foreign countries for critical applications (e.g., government websites), or connections to servers in unexpected locations.
A security analysis published in 2023 discusses malicious JavaScript code that queried an Ethereum provider to retrieve a second-stage domain from a malicious smart contract, which resolved to IP addresses located in a foreign country~\cite{guardio2023etherhiding}.
In another security incident report in 2023, skimmers targeted sites across the globe and exfiltrated data to a server located in Japan~\cite{blackberry2023silentskimmer}.

\Cref{app:blocks}~\Cref{lst:policy_location} encodes a broad check for resources hosted by servers in unexpected locations, by performing a DNS query for the domain, and geolocating the resulting IP address of the remote server.
Using this result, \aliasSystemName{} can compare the country of the result against a pre-defined allowlist or blocklist, and compare the distance to the remote server against a distance threshold as a means to enforce a maximum distance restriction.
Variations of this policy can additionally check for suspicious indicators based on the DNS responses to queries for the requested domain, such as the number of distinct IP addresses or countries~\cite{bilge2011exposure} or autonomous systems for those IP addresses~\cite{antonakakis2010building}.

\subsection{Policy: CMS Core File Integrity}
Content management systems (CMS) comprise a significant fraction of the websites deployed today --- some sources report that over 75\% of the top 1M sites by traffic are built with CMS software~\cite{builtwith2024cms}.
Given their prevalence, attackers frequently target sites that use CMS software, and one of the common targets are the ``core'' files that are sent to users (e.g., WordPress jQuery~\cite{sucuri2022wpjstoads} and themes~\cite{sucuri2025corewptheme}), or encode server-side logic (e.g., session management in Magento~\cite{sucuri2021ccfile}).

\Cref{app:blocks}~\Cref{lst:policy_core_files} encodes core file checks as a special type of \aliasSystemName{} policy that can be run to check for both client-side (e.g., JavaScript) or server-side (e.g., PHP) core file integrity.
Client-side checks would be performed as usual, but for server-side checks, the verifier would require access to the corresponding source files used by the web application.
For these cases, the policy would need to reference another source to obtain the expected content for core files, and compare it to the content that is actually received by the client, or the content that is actually present in the web application, respectively.
This is in contrast to SRI and CSP, which cannot help with server-side checks, but can help perform client-side checks, provided the CMS software supports their use.
Further, content checks with SRI or CSP do not apply to all file types that may be considered core CMS files.
\graphicspath{ {./figs/04_eval/} }

\section{Evaluation}
\label{sec:04_eval}

To evaluate the performance overhead of our prototype, we implemented a pipeline that simulated deployments by injecting \aliasSystemName{} service worker registration scripts into responses from the server.
In this section, we evaluate the performance overhead of our \aliasSystemName{} prototype, and evaluate the robustness of several proposed policy building blocks from \Cref{sec:03_blocks}.
As the simulated deployment methodology is not a critical component of \aliasSystemName{}, we refer readers interested in the technical details to \Cref{app:simulated_deployment}, which discusses them at depth.


\subsection{Domain Sample}
\label{subsec:04_sample}

We selected a representative sample of domains to be used in our performance and policy evaluations. 
First, we obtained a ranked list of the top one million domains from Tranco on \paramTrancoListDate{}.
We divided the domains into high-, mid-, and low-ranking buckets of increasing size, [1,~1K], (1K, 100K], and (100K, 1M) respectively.
We then randomly sampled \paramTrancoNumPerBucket{} domains from each bucket.
The resulting sample contains domains that span a diverse array of services, as outlined in \Cref{app:domain_sample_categorization}, \Cref{tab:domain_categories}.
We filtered the selected domains, excluding \numDomainsExcludedByKeyword{} domains with adult content keywords, \numDomainsNotConnectable{} domains for which our crawler did not receive a response in 30 seconds, and \numDomainsWithExistingSW{} domains with existing service workers.
The remaining \numPrefilteredDomains{} domains were used in our performance evaluation, following the simulated deployment methodology described in \Cref{app:simulated_deployment}.

\paragraph{Request Patterns.}
Before discussing the evaluations in detail, we characterize the inclusion patterns of the domains in our sample from the perspective of a \aliasSystemName{} administrator by using the generated data from our simulated deployments.
For each domain, the median number of external URLs that were requested is 197, number of external origins is 7, and number of URLs per external origin is 32.
As expected, third-party advertising and tracking sites are requested very often, from many of the domains in our sample.
We observe that \code{googletagmanager.com} is the most commonly-requested external origin, with 170 domains in our sample triggering an average of 2 unique URLs to this external origin on each visit, with a total of 1,247 different URLs encountered throughout our performance evaluation.
This site is responsible for delivering advertising-related scripts, based on the URL parameter \codemono{id} that identifies the set of scripts to load. 
In contrast, we also observed external origins that were used by only one domain in our sample, but were requested with many unique URLs per visit.
We found that, on each visit, \code{whatsapp.com} requested, on average, 63 unique transient URLs from \code{whatsapp.net}, most of which were images that return an error message outside a regular browsing session (e.g., ``Bad URL timestamp/hash'').
Similarly, \code{sunsky-online.com} triggered requests for 20 URLs to \code{myipadbox.com} on average, but these were static URLs that corresponded to images of their e-commerce inventory.

\paragraph{Takeaways.}
Domains in our sample are representative in terms of ranking and categorization: our sampling methodology is designed to represent low-, middle-, and high-ranked domains, and also to select domains that serve a diverse array of functions.
There is no universally-applicable third-party resource inclusion pattern, and we observe evidence of different request inclusion patterns in our simulated deployments.
As such, administrators must first understand their existing third-party resources before writing policies.
The multi-stage deployment strategy of \aliasSystemName{} is expressly suited to assist administrators in developing an understanding of existing resources by allowing administrators to recognize URL patterns and develop policies.
Furthermore, it may be valuable to develop common policies for commonly-included third parties; for example, policies designed for Google Tag Manager may be attractive to site administrators and ease adoption of systems like \aliasSystemName{}. 

\subsection{Performance Overhead}
\label{subsec:04_perf}

\begin{table}[t]
    \centering
    \caption{
        Different stages of the performance overhead evaluation imposed by our prototype \aliasSystemName{}.
    }
    \resizebox{0.7\columnwidth}{!}{
        \begin{tabular}{l|c|p{5cm}}
            \toprule
            \textbf{Stage}       &
            \textbf{Overhead Measurement}
            \\
            \midrule
            \fmtEvalStageZero{}  &
            Baseline measurement with no SW
            \\
            \fmtEvalStageOne{}   &
            No-op request interception with SW
            \\
            \fmtEvalStageTwo{} &
            Queries link status from no-op API
            \\
            \fmtEvalStageThree{}  &
            Queries link status from normal API
            \\
            \bottomrule
        \end{tabular}
    }
    \label{tab:eval_modes}
\end{table}
\begin{figure}[t]
    \centering
    \includegraphics[width=0.9\columnwidth]{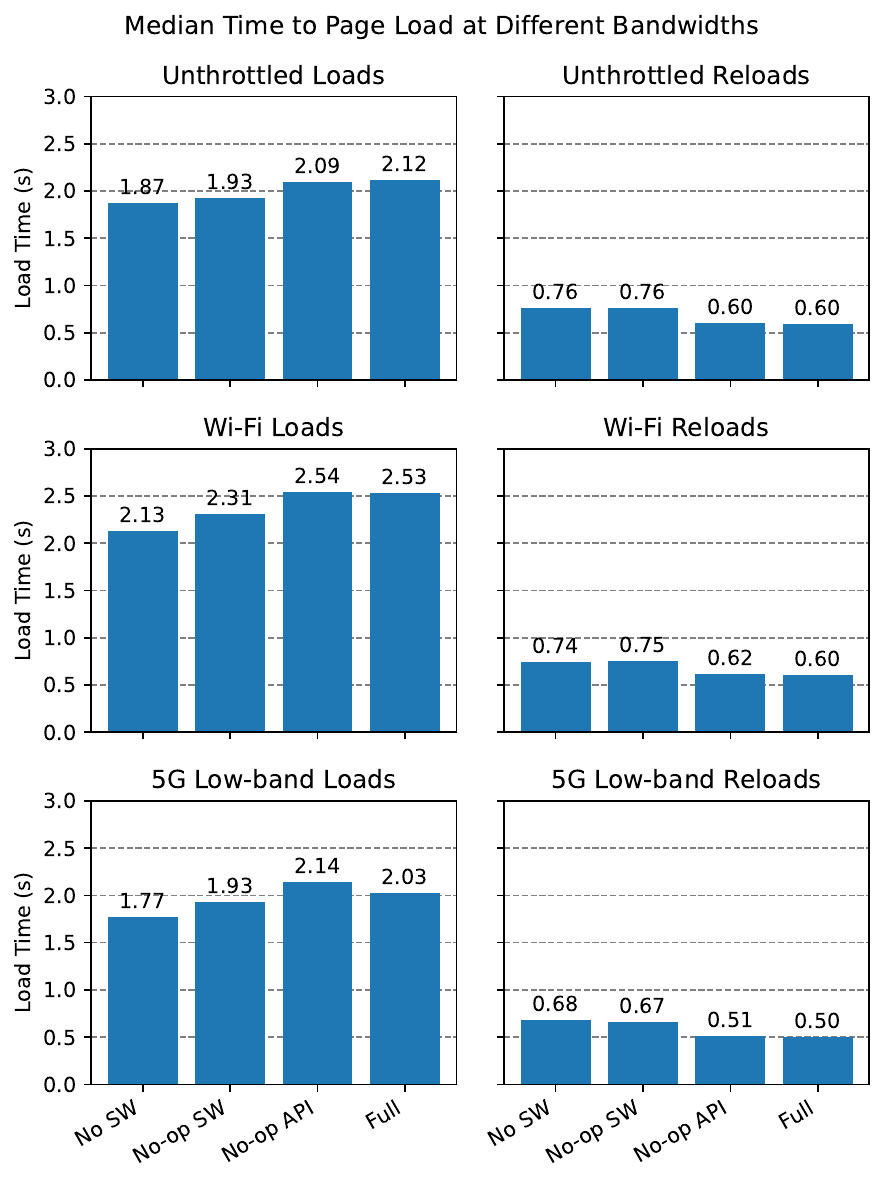}
    \caption{
        Bar charts depicting the total overhead for users introduced by the \aliasSystemName{} prototype implementation.
    }
    \label{fig:bar_perf}
\end{figure}

We conducted our performance evaluation on the \numPrefilteredDomains{} domains by simulating the deployment of \aliasSystemName{} as in \Cref{app:simulated_deployment}, which injects service worker registration scripts in server responses.
Next, we measured the overall page load times while progressively activating more components of \aliasSystemName{} to analyze the overhead contributed by each one, under varying network speeds.
\Cref{tab:eval_modes} summarizes the different stages of this incremental activation process, starting with a baseline measurement with no SW, and progressing to a full activation with normal \aliasSystemName{} functionality.

The overhead measurements are reported as median page load times in a series of bar charts in \Cref{fig:bar_perf}.
Each of the plots represents a different network configuration, with the left half denoting the median page load times on the first visit, and the right half denoting those for subsequent (second) visits.
Each domain is visited \paramNumEvalOverheadTrials{} times for each combination of the network configuration and the evaluation stage, and the median of these is taken to be the page load time for that domain under those conditions.
From the \numPrefilteredDomains{} domains after prefiltering, an additional \numDomainsMissingStage{} were filtered out because they were missing data for at least one evaluation stage, which were caused by transient failures of the evaluation infrastructure (e.g., unresponsive browsers on evaluation workers) and transient network issues (e.g., timeouts when connecting to the website).
The height of each bar in a plot corresponds to the median of the page load times across all remaining \numRemainingDomains{} domains for the corresponding evaluation stage.
For example, in the 5G low-band network profile, the median page load time across the \numRemainingDomains{} domains is 1.77 seconds in the \fmtEvalStageZero{} stage.

If we instead consider the 90th percentile of load time overhead, the resulting plot is very similar to \Cref{fig:bar_perf}, except that times are a few hundred milliseconds larger. 
At the 99th percentile, the same is true, but with greater overhead for load times that take over $2\times$ compared to the 50th percentile of loads (e.g., WiFi has a 2.13s baseline and 2.53s total at the \nth{50} percentile, compared to a 4.92s baseline and 6.26s total at the \nth{99} percentile).
Although we did not measure the execution time for each policy verification, proactive re-execution of policy checks can warm the verification cache, preventing user requests from triggering expired policy verifications and improving user experience as mentioned in \Cref{subsec:02_backend}.

\paragraph{Takeaways.}
In each of the three network configurations, we observe that the overhead introduced at the last two evaluation stages are significantly higher than the overhead introduced at the \fmtEvalStageOne{} stage.
This aligns with expectations: \fmtEvalStageOne{} does not introduce any additional network communication, whereas the \fmtEvalStageTwo{} and \fmtEvalStageThree{} stages do on the first load of a page.
The overhead introduced by a no-op service worker that intercepts requests, and immediately returns from the intercept handler, is \numOverheadUnthrottledNoopSW{} ms, \numOverheadWiFiNoopSW{} ms, and \numOverheadFiveGNoopSW{} ms for the three network profiles.
In contrast, the overall, additional overhead to the median first load time is \numOverheadUnthrottledMilliseconds{} ms, \numOverheadWiFiMilliseconds{} ms, and \numOverheadFiveGLowBandMilliseconds{} ms for the unthrottled, Wi-Fi, and 5G low-band speeds respectively.
Despite this, the additional overhead is within the generally-accepted range for performant UI response times~\cite{jakob1993response}.

In contrast to the first page load times, the median page reload times are unusual: the median page reload times for the last two stages are lower than those for the first two.
On the second load, the \fmtEvalStageTwo{} and \fmtEvalStageThree{} service worker does not perform any additional network requests: it will consult its local cache and allow the request to pass through.
Thus, we hypothesize that this phenomenon is caused by subtle differences in browser behavior when a service worker's request interception logic is not empty, as this phenomenon does not manifest for the \fmtEvalStageOne{} reload measurements.
In short, the actual overhead for reloading pages for \fmtEvalStageTwo{} and \fmtEvalStageThree{} should be similar to \fmtEvalStageOne{}, as the only additional work introduced is the SW checking its local cache.

In summary, intercepting requests with a service worker introduced \numOverheadUnthrottledNoopSW{} ms, \numOverheadWiFiNoopSW{} ms, and \numOverheadFiveGNoopSW{} ms total overhead to the median page first load time across the three different network profiles.
Full operation introduced \numOverheadUnthrottledMilliseconds{} ms, \numOverheadWiFiMilliseconds{} ms, and \numOverheadFiveGLowBandMilliseconds{} ms overhead to the median page first load time for the unthrottled, Wi-Fi, and 5G low-band speeds respectively.
Additionally, the overhead on subsequent visits with policies that do not need to be re-verified is negligible.

\subsection{Policy Evaluation}
\aliasSystemName{} also offers a unique vantage point for administrators to analyze their site: even without custom policies, default policies (\S\ref{sec:03_blocks}) can present new insights into the dependency usage of a web application.
We discuss several ways that administrators, who may not have the expertise to write customized and robust integrity policies, can use \aliasSystemName{}.
In particular, we demonstrate how \aliasSystemName{} can provide insights into dependency usage and generate thresholds for several default policies for a site.
The following analyses are performed by components of the open-sourced artifact accompanying this text.

\subsubsection{Data Source}
To quantify this discussion, we leverage the data archived by the Common Crawl (CC) project~\cite{commoncrawl} to emulate longitudinal analyses.
The CC project periodically archives the web with their crawlers and provides its data for public use in an \emph{index} defined by the year and the week in which the crawl was finished (e.g., \code{2024-42} refers to week 42 in the year 2024).
We used \paramCCNumContiguousIndexes{} contiguous indices spanning approximately \paramCCTimeWindowWeeks{}~weeks from the index \paramCCStartIndex{} to \paramCCEndIndex{}, in order to analyze the links present on domains in a longitudinal manner.

For each index, we attempted to download the corresponding snapshot of the landing page of each domain in the sample from \Cref{subsec:04_perf}, and extracted all the links in the HTML of the landing page.
We filtered this data so that we only consider the \numCCDomains{} domains that were successfully crawled and archived in all \paramCCNumContiguousIndexes{} snapshots.
Various factors contribute to this low number, with the most prominent being that sites may explicitly forbid the CC bots from crawling their site, present an HTTP 301 response that redirects to another URL that was not successfully archived (e.g., \code{example.com} redirects to the URL \code{www.example.com} which was not archived), or were not online or crawled at the time.
Regardless, the remaining \numCCDomains{} constitute a reasonable sample for this discussion.

\begin{figure}[t]
    \centering
    \includegraphics[width=1.0\columnwidth]{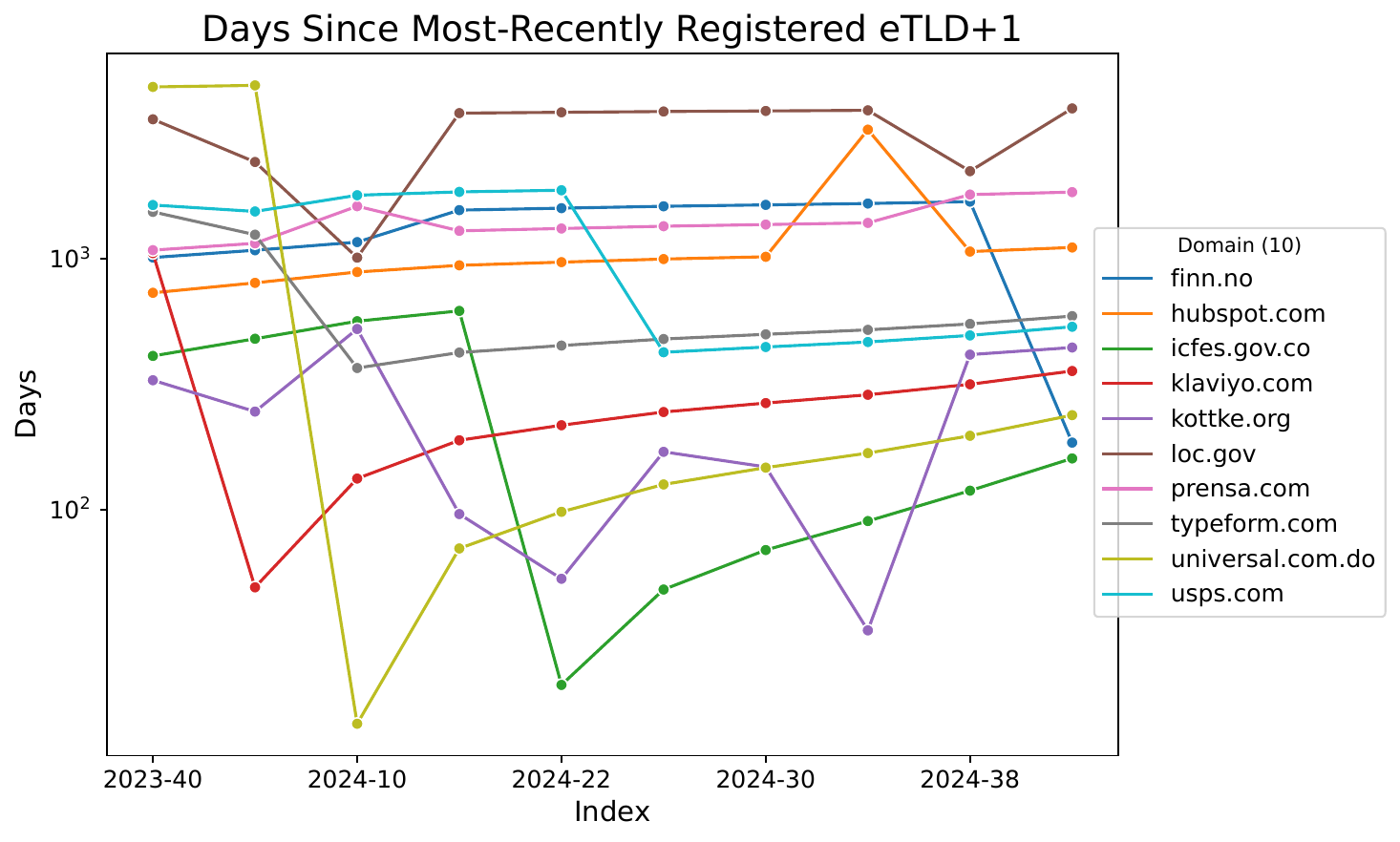}
    \caption{
        Number of days since the most recent registration of all linked eTLD+1 domains.
    }
    \label{fig:line_days_reg}
\end{figure}
\subsubsection{Policy: Domain Lifecycle}
\Cref{fig:line_days_reg} depicts the thresholds that would be necessary to use the policy described in \Cref{subsec:04_policy_domain_lifecycle} on a subset of the \numCCDomains{} domains from the CC dataset.
In particular, for each domain at every CC index, we aggregated the eTLD+1 of all linked domains on its landing page, and used a commercial passive DNS database~\cite{farsightdnsdb} to identify the approximate date when each eTLD+1 domain was registered by using similar methodology as other studies for passive DNS analyses~\cite{so2025lost}.
Then, we approximated the date corresponding to the CC index by taking the first day of the specified week, with weeks corresponding to the ISO 8601 definition~\cite{iso8601}.
Afterwards, for each of the \numCCDomains{} domains, we computed the number of days since the registration of the most recently registered eTLD+1 among its links.
\numCCLifecycleStableDomains{} of the domains exhibited monotonically increasing behavior, indicating they never included a newer eTLD+1.

We observe two main patterns in \Cref{fig:line_days_reg}, which plots a sample of the \numCCLifecycleUnstableDomains{} remaining domains.
The first is that there is a general monotonically-increasing trend for each domain, indicating that domains do not continuously add recently-registered domains to their supply chain.
This supports the use of the corresponding policy discussed in \Cref{subsec:04_policy_domain_lifecycle}, which restricts the use of newly-registered domains.
There is one notable exception displayed in the plot: \nolinkurl{kottke.org}.
This domain hosts one of the oldest blogs on the web~\cite{kottkeorg} and by its nature, includes links to a multitude of other domains on its landing page, resulting in erratic fluctuations with no discernible trend.
The second pattern is that most domains, after including a domain that was more recently registered than all other domains in their supply chains, continue with the general monotonically-increasing trend.
This suggests that the domains only periodically include such new domains, thereby implying that if the domain lifecycle policy were deployed, it does not require frequent updates to the threshold.

In fact, the threshold for the duration of time since registration of linked eTLD+1 domains can be automatically and incrementally raised, making it more difficult for adversaries to introduce new domains.
If administrators desire to add a new domain, they can temporarily lower the threshold, before continuing to harden the policy threshold.
We observe that most domains linked to newer eTLD+1s when introducing new services (e.g., \nolinkurl{usps.com} linking to \nolinkurl{uspssmartpackagelockers.com} 424 days after its registration and \nolinkurl{universal.com.do} linking to \nolinkurl{asistenciauniversal.om.do} only 14 days after its registration), or temporal content (e.g., \nolinkurl{loc.gov} linking to \nolinkurl{blackhistorymonth.gov} in February, approximately 1,011 days after its registration).
In the former case, these domains generally keep these services, and this behavior will manifest as a sudden drop and gradual increase in the plot.
In the latter case, these domains generally remove these linked eTLD+1s after a certain time window, and this behavior will manifest as a sudden drop, a sudden increase to the same level before the drop, and then continue to gradually increase.

\begin{figure}[t]
    \centering
    \includegraphics[width=1.0\columnwidth]{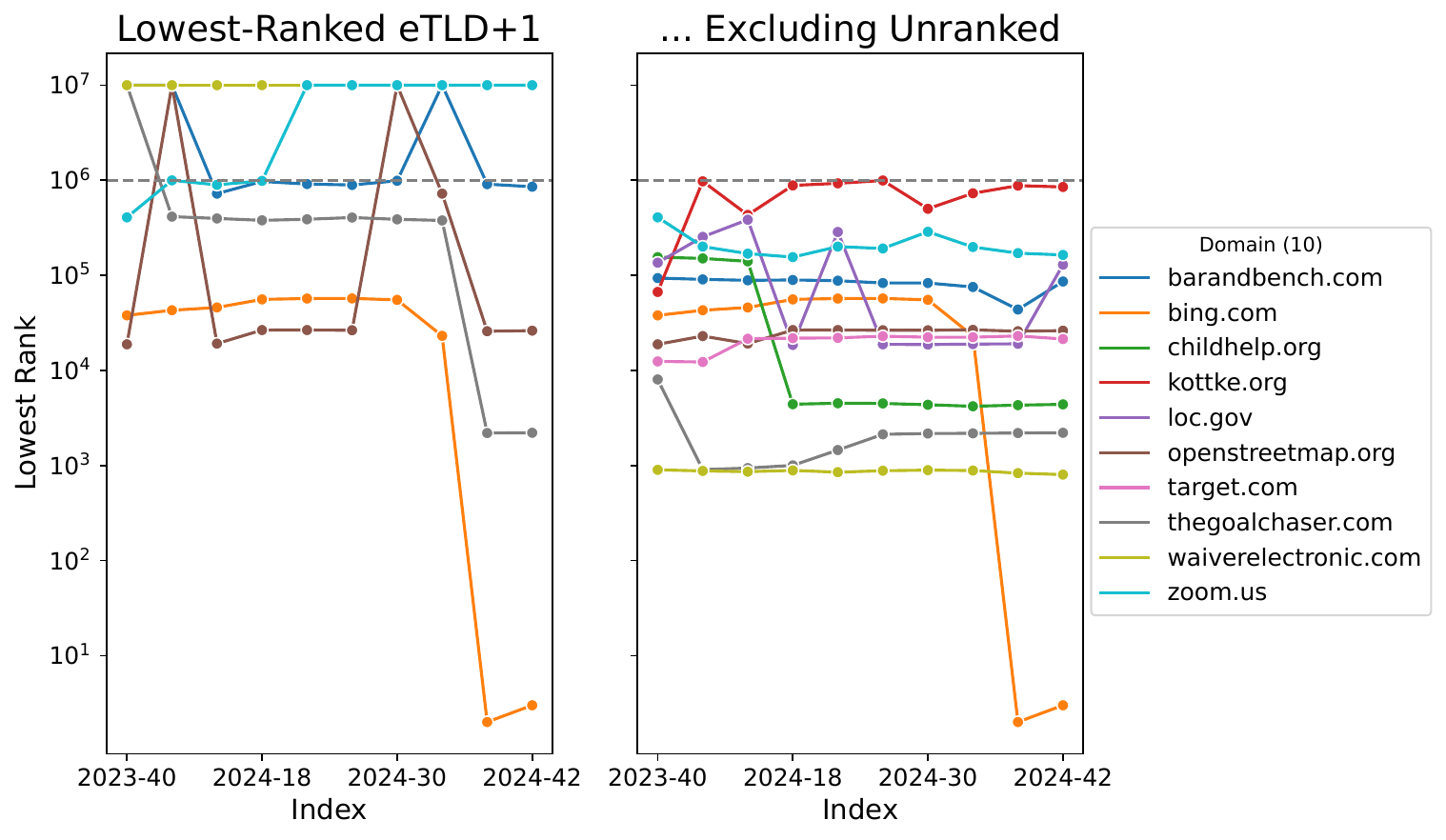}
    \caption{
        The lowest-ranked domains of all linked eTLD+1 domains.
        The left plot imputes a default rank for unranked domains, whereas the right excludes them.
        The dashed line represents the Tranco top 1M.
    }
    \label{fig:line_lowest_ranked}
\end{figure}
\subsubsection{Policy: Domain Ranking}
We conduct a similar analysis to examine the feasibility of the policy discussed in \Cref{subsec:04_policy_domain_ranking}.
Using the approximate dates for each CC index as previously described, we download the Tranco top 1M ranking list generated on those days, and look up the ranking for each of the eTLD+1 domains that were linked on the landing pages of the \numCCDomains{} CC domains.
\Cref{fig:line_lowest_ranked} depicts the results for a sample of the \numCCLowestRankUnstableDomains{} domains that experienced significant changes in their trends, through two perspectives: the left plot, which includes unranked domains by imputing a default rank value of $10^7$, and the right, which excludes them.

We observe several patterns in \Cref{fig:line_lowest_ranked} that are similar to those in \Cref{fig:line_days_reg}.
The first is that the exclusion of unranked domains yields an approximately constant trend for domains that do not introduce new, lower-ranked eTLD+1s.
As domains may experience slight changes from one day to the next, the trends will appear approximately constant with slight fluctuations.
The second pattern is that when domains introduce domains for new services, these are usually unranked as the domains themselves are new, but may become ranked if they are kept for extended periods of time (e.g., \nolinkurl{childhelp.org} linked to \nolinkurl{childhelphotline.org} since \code{2023-40}, and it was finally ranked 995K on \code{2024-33}).
If the lowest-ranked is instead removed, then it will present the opportunity to significantly harden the ranking threshold (e.g., \nolinkurl{bing.com} which removed \nolinkurl{takelessons.com} after \code{2024-30} and \nolinkurl{start.gg} after \code{2024-33}).
Alternatively, if the new linked eTLD+1s correspond to temporal content, they may be removed from the main site before they appear on ranking lists (e.g., \nolinkurl{loc.gov} linked to the unranked \nolinkurl{jewishheritagemonth.gov} on only \code{2024-22}).
Regardless of the scenario, administrators can manage the domain ranking policy similarly to the domain lifecycle policy.
When new domains are added or removed, the policy threshold, or policy-specific exceptions, can be adjusted, before continuing to gradually harden the threshold.

\subsubsection{Summary}
Overall, the archived CC snapshots of these domains support the feasibility of the domain lifecycle and domain ranking policies from \Cref{sec:03_blocks}.
Although we are unable to evaluate the remaining policy building blocks with CC data, or evaluate how well they detect new attacks, we argue that integrity policies in \aliasSystemName{} can provide customizable integrity guarantees in flexible manners, and there exist general policies (e.g., the domain lifecycle and domain ranking) that add valuable protections for users that existing integrity mechanisms cannot.
\section{Related Work}
\label{sec:05_related}
To our knowledge, there exists no other work that introduces the concept of granular and flexible integrity policies and an automatic enforcement system that can provides integrity guarantees for clients by blocking network requests to resources whose configured policies are violated.
Prior works have introduced the notion of integrity to resources, such as newer versions of Content Security Policy (CSP) and Subresource Integrity (SRI), but the goals, flexibility, and design of such methods are significantly different.
This paper marks the first attempt in extending the traditional notion of data integrity in flexible ways that are more fitting for the modern web, enabling integrity to be applied to different dimensions.
Further, we provide open source prototype implementations, and evaluate them, as part of this work to hopefully foster more research in this area.
In the rest of this section, we discuss prior works that introduce link integrity guarantees, security policies, and malicious JavaScript detection.

\paragraph{Content Security Policy}
CSP is the most closely related work to the system proposed in this paper.
Originally proposed to defend against Cross-Site Scripting (XSS) attacks~\cite{stamm2010reining}, it adds an HTTP header that can be included by servers to define allowlists of trusted origins for different resource types, such as scripts, media, and images, on individual pages.
Browsers must check all network requests to make sure that the origin of the request URL is included in applicable policies, if specified.
If violations are found, browsers generate and send a report to a preconfigured server in the policy header, and can block the request if the policy is configured in the enforcement mode.
In summary, CSP can be used to \emph{restrict the resources that are loaded on individual pages to predefined sets of trusted origins}.
Furthermore, extensions to its standard have enabled mechanisms to verify the authenticity of inline scripts via nonces and content hashes~\cite{w3c2024csp3}.

The first major difference between the two systems --- besides the fact that policies are verified by the browser in CSP, and by a client service worker in \aliasSystemName{} --- is the threat model, and subsequently what is considered to be \emph{trusted}.
CSP originally considers a resource to be trusted if and only if the resource is loaded from an explicit allowlist of origins.
In short, \emph{CSP trusts certain origins, or exact file contents}.
\aliasSystemName{} is concerned with defending against supply chain attacks.
The concept of integrity is much more flexible and is reflected in the generalizability of integrity policies.
In short, \emph{\aliasSystemName{} trusts in more dimensions than the origin of the provider and exact file contents, including resource behavior, request context, and location information.}
Thus, the set of integrity dimensions afforded by CSP is a strict subset of that of \aliasSystemName{}, but this does not mean that \aliasSystemName{} provides the same set of e.g., XSS protections from CSP.

The other major difference relates to \emph{when and where policies are verified}.
In CSP, policies are verified by the browser which has received the set of policies for the current page in the HTTP headers.
In \aliasSystemName{}, policies are verified on demand by dedicated server-side workers that cache decisions for all users.
\emph{Separating the component that performs the actual verification from the browser enables \aliasSystemName{} to use complex policies} and cache its verification decisions for reuse among all users, minimizing the overhead imparted to any individual user.

\paragraph{Subresource Integrity}
SRI is another existing standard that provides integrity guarantees of web resources.
In contrast to CSP, SRI is expressly designed to combat this problem, and it does so by defining a new attribute in the HTML script tag that specifies the expected hash value(s) for the content of the script~\cite{w3c2016sri}.
This enables user agents to compare the hash value of the \emph{received} content against the hash value of the content that is \emph{expected}, loading received content if and only if the hash (if specified) matches.

In the same vein as CSP integrity, such guarantees are \emph{strict} by nature and do not allow any type of change in content, regardless of the scope of the change.
Prior work has empirically shown that strict integrity is not a good fit to ensure the integrity of JavaScript, one of the primary resource types on the web.
Steffens et al. found that high-profile parties randomize select parts of their scripts, rendering it impossible to apply SRI for them~\cite{steffens2021block}.
Similarly, So et al. reported that even if every static script were protected with SRI, it would not provide the intended security guarantees because sites often retrieve both static and non-static scripts from third-party origins~\cite{so2023more}.

The lack of adoption of SRI seems to agree with these studies that challenge the usability and practicality of strict integrity verification.
SRI was first proposed as a standard in 2016 by the World Wide Web Consortium~\cite{w3c2016sri}, and multiple studies conducted over the years have found the adoption rate to be low.
Kumar et al.~\cite{kumar2017security} and Shah and Patil~\cite{shah2018measurement} both found that less than 1\% of sites use SRI in 2017 and in 2018 respectively.
Chapuis et al.~\cite{chapuis2020empirical} reported that the adoption had increased to 3.4\% of all webpages in 2020 in a longitudinal study with a much larger sample size.

\paragraph{Security Policies}
There are a number of prior works that introduce security policies for the web, aside from CSP, to guarantee that JavaScript execution is limited to trusted code.
One of the first proposals by Jim et al. to defend against cross-site scripting, Browser-Enforced Embedded Policies (BEEP), introduced the notion of website-defined security policies to be used by the browser to determine which scripts are allowed to run~\cite{jim2007defeating}.
Another proposal by Oda et al. reimagined web resource usage permissions if cross-origin communications were required to be mutually approved, finding that it would defend against otherwise-successful cross-site scripting (XSS) and request forgery (XSRF) attacks~\cite{oda2008soma}.
Other works proposed policy systems that acted directly on JavaScript.
Reis et al. proposed Browsershield, a system that intercepts and rewrites JavaScript code subject to site-defined execution policies, aims to avoid executing code that exploits web browser vulnerabilities~\cite{reis2007browsershield}.
Similarly, Meyerovich and Livshits proposed Conscript to grant browsers the ability to enforce fine-grained security policies for JavaScript, effectively adding constraints to the code during execution~\cite{meyerovich2010conscript}.
Phung et al. proposed FlashJaX, a cross-language inline reference monitor that enforced security policies on third-party, mixed JavaScript and ActionScript content~\cite{phung2014between}, which also used a robust client-side mechanism that did not require browser modifications.

When these designs were proposed more than a decade ago, the primary security incidents were XSS, XSRF, and web browser vulnerabilities, and their considered threat models accordingly reflect this.
As such, these systems do not immediately align in the threat model of supply chain attacks where users may receive unexpected or malicious content from trusted origins.

\paragraph{JavaScript Integrity}
A different line of related work studies the feasibility of different integrity schemes for JavaScript in terms of producing signatures and fingerprints with program analysis.
\aliasSystemName{} is a system designed to offer flexible integrity policies, and an ideal policy would compare the signature or fingerprint of a script against a configured allowlist, using a signature or fingerprint scheme that is robust even in the presence of content changes.

It has proven difficult to produce such a robust, general signature scheme.
Strict integrity schemes such as those created by Nakhaei et al.~\cite{nakhaei2020jssignature} and Mignerey et al.~\cite{mignerey2020ensuring} do not address the threat model or require adding a new component to the web public key infrastructure.
Soni et al.~\cite{soni2015sicilian} and Mitropoulos et al.~\cite{mitropoulos2016train} offered novel, relaxed integrity schemes to generate structural signatures and contextual script fingerprints respectively.
However, a recent study by So et al. found that these relaxed integrity schemes are unstable in the context of modern web scripts~\cite{so2023more}.
\section{Discussion}
\label{sec:06_disc}
This text introduces the concept of integrity policies, an application-agnostic design of a corresponding verification and enforcement system to prevent supply chain attacks, and basic, yet efficient, policies that can be readily implemented and enforced based on commonalities from recent security incidents.
Furthermore, we evaluate the overall overhead introduced by our prototype \aliasSystemName{} in the form of page load times for simulated deployments, and evaluate several proposed policies, finding that our prototype \aliasSystemName{} introduced minimal overhead, and that policies can serve as building blocks.

\subsection{Limitations}
Despite the advantages, there are inherent limitations of our design: \aliasSystemName{} will not be able to intercept WebSocket traffic.
Additionally, the \aliasSystemName{} client will also be susceptible to any exploits that leverage the service worker design and implementation (e.g., privacy sniffing~\cite{karami2021awakening}).
Also, server-side integrity policy verification introduces a non-trivial problem: server-side workers are verifying that remote resources are safe on behalf of clients.
There is no guarantee that the server-side workers will receive the same responses as clients.
Thus, it may be the case that cloaking of a compromised resource may be able to bypass an integrity policy.
Further, policies that are robust in the face of content updates may be difficult to write, but there may be better-fitting policies that check the integrity of other dimensions for such resources.
In addition, the evaluation of our prototype did not take into consideration the execution time of policies, but the overhead is negligible when verifications are performed periodically by verifiers that do not block the request flow.
Our evaluation also did not take into consideration the latency between the client and the \aliasSystemName{} server, but techniques such as load-balancing and distribution of hosting servers are expressly designed to tackle this problem.

Additionally, as with all deployed applications, a deployment of \aliasSystemName{} increases the attack surface as the number of nodes in the organization's infrastructure itself will increase.
In addition, policy writers will need to cautiously use external libraries, as it may be possible to induce a supply chain attack on a \aliasSystemName{} verifier that relies on external code.
However, we remark that the addition of \aliasSystemName{} does not introduce any additional, significant attack surface to website visitors.
An attacker that can maliciously modify the \aliasSystemName{} client service worker itself would also be capable of modifying any of the other first-party content, or injecting their own malicious service worker. 

\subsection{Design Considerations}
The prototype \aliasSystemName{} which we introduce in this work has a number of different aspects that can be adjusted.
One area is the choice of server-side policy verification --- an alternate implementation can include client-side verification for non-content-related policies (because service workers cannot access the contents of cross-origin responses), or directly enable support for content-related policies by implementing the enforcement logic in the browser.
This is beneficial because client-side verification guarantees that policies will be verified on the responses that clients receive.
By delegating verification to a server-side component, policies must avoid depending on user-specific information in resource requests.
However, as previously mentioned, client-side verification may not be ideal as we suspect that it will introduce significant overhead for users, and the current implementation is sufficient for a working prototype.

Another area that can be optimized is the verification protocol between the client SW and the API server.
As each newly-observed resource triggers an additional network request in the prototype, it is likely that minimizing the overhead of communications can drastically minimize the overall delay.
One method to achieve this is to modify the API server to eagerly respond to navigation requests with the status of resources that are expected to load on the first-party page, or to move the client SW and the API server communication to a WebSocket channel to reduce the overhead of TLS handshakes.

Lastly, additional features can be readily incorporated into \aliasSystemName{}.
One feature involves leveraging push subscriptions to send notifications from the server to users, granting the ability to force cache refreshes upon failed or revoked verifications, and also remove the need for the client service worker to regularly poll the API server.
Another feature that would prove to be useful is to add support to track the contents of resources to enable longitudinal analyses of content changes for administrators to review when, and how, resources change.

\section{Conclusion}
\label{sec:07_conc}

In this paper, we introduced the concept of granular and flexible integrity policies that declare the (un)expected properties of resources in different dimensions, and the application-agnostic design of a corresponding verification and enforcement system \aliasSystemName{}.
In the design of \aliasSystemName{}, policy verification is performed by dedicated server-side workers who cache decisions for reuse among all users, and policy enforcement is upheld in the form of a service worker that is installed in user browsers.
We also introduce universally-applicable integrity policies to serve as building blocks for a comprehensive set of integrity policies and discuss how they could have prevented recent supply chain attacks reported in the industry.
Finally, we implemented an open-source prototype of \aliasSystemName{} and found that it adds minimal performance overhead and no loss of functionality to first-party applications during a systematic evaluation of the overhead introduced by the different components.
The overall overhead encountered during the first load of a page is several hundred milliseconds, and there is negligible overhead during subsequent loads, for each of the tested network configurations.
We also examine the feasibility of several proposed policy building blocks, finding that they suit the dependency usage patterns of sites and would incur minimal overhead for administrators.

\begin{acks}
We thank the reviewers for their helpful comments.  This work was supported by the Office of Naval Research (ONR) under grant N00014-24-1-2193 as well as by the National Science Foundation (NSF) under grants CNS-1941617, CNS-2126654, and CNS-2211575.
\end{acks}


\bibliographystyle{ACM-Reference-Format}
\balance
\bibliography{references}

\appendix
\section{High-Level Policy Building Blocks}
\label{app:blocks}
This section contains a series of algorithms that correspond to the proposed policy building blocks described in \Cref{sec:03_blocks}.
In particular, \Cref{lst:policy_reputation} corresponds to \Cref{subsec:04_policy_domain_ranking}; \Cref{lst:policy_threat_intel} corresponds to \Cref{subsec:04_policy_threat_intel}; \Cref{lst:policy_dependencies} corresponds to \Cref{subsec:04_policy_dependencies}; \Cref{lst:policy_visibility} corresponds to \Cref{subsec:04_policy_sri_violations}
They refrain from technical details as they are not critical to the goals of this paper, which introduces \aliasSystemName{}.
However, interested readers can find their implementations in the prototype that will be in the accompanying artifact submission.

\begin{algorithm}[h]
    \small
    \caption{
        A sample integrity policy that denies requests to low-ranking domains, unless explicitly allowlisted.
    }
    \label{lst:policy_reputation}
    \begin{algorithmic}[1]
        \State \textbf{Class} {PolicyLowRanked}
        \State \ \ \ \ \textbf{Variable} locationSource = "example.com/.*"
        \State \ \ \ \ \textbf{Variable} locationTarget = ".*"
        \State \ \ \ \ \textbf{Variable} logic = IsLowRanked
        \State \ \ \ \ \textbf{Variable} output = False
        \\

        \Procedure{IsLowRanked}{req}
        \State thresh $\gets$ GetRankingThreshold()
        \State lowRanked $\gets$ IsLowRanked(req.domain, thresh)
        \State allowlisted $\gets$ IsAllowListedForRanking(req.domain)
        \State \Return lowRanked AND NOT allowlisted
        \EndProcedure
    \end{algorithmic}
\end{algorithm}

\begin{algorithm}[h]
    \small
    \caption{
        A sample integrity policy that denies requests for resources that are flagged by threat intelligence services or displays suspicious behavior such as camouflaging or cloaking.
    }
    \label{lst:policy_threat_intel}
    \begin{algorithmic}[1]
        \State \textbf{Class} PolicyThreatIntelligence
        \State \ \ \ \ \textbf{Variable} locationSource = "example.com/.*"
        \State \ \ \ \ \textbf{Variable} locationTarget = ".*"
        \State \ \ \ \ \textbf{Variable} logic = IsFlaggedByThreatIntel
        \State \ \ \ \ \textbf{Variable} output = False
        \\

        \Procedure{IsFlaggedByThreatIntel}{req}
        \State isFlagged $\gets$ CheckThreatIntel(req)
        \State usesCamouflaging $\gets$ RunCamouflagingDetection(req)
        \State \Return isFlagged OR usesCamouflaging
        \EndProcedure
    \end{algorithmic}
\end{algorithm}

\begin{algorithm}[h]
    \small
    \caption{
        A sample integrity policy that denies requests for resources whose dependencies have changed.
    }
    \label{lst:policy_dependencies}
    \begin{algorithmic}[1]
        \State \textbf{Class} PolicyChangedDependencies
        \State \ \ \ \ \textbf{Variable} locationSource = "example.com/.*"
        \State \ \ \ \ \textbf{Variable} locationTarget = ".*"
        \State \ \ \ \ \textbf{Variable} logic = HasChangedDependencies
        \State \ \ \ \ \textbf{Variable} output = False
        \\

        \Procedure{HasChangedDependencies}{req}
        \State cached $\gets$ GetPreviousDependencies(req)
        \State curr $\gets$ GetCurrentDependencies(req)
        \State \Return cached == curr
        \EndProcedure
    \end{algorithmic}
\end{algorithm}

\begin{algorithm}[h]
    \small
    \caption{
        A sample integrity policy that denies requests for resources with TLS connection errors or SRI violations.
    }
    \label{lst:policy_visibility}
    \begin{algorithmic}[1]
        \State \textbf{Class} PolicySRIViolations
        \State \ \ \ \ \textbf{Variable} locationSource = "example.com/.*"
        \State \ \ \ \ \textbf{Variable} locationTarget = ".*"
        \State \ \ \ \ \textbf{Variable} logic = FailsSRICheck
        \State \ \ \ \ \textbf{Variable} output = False
        \\

        \Procedure{FailsSRICheck}{req}
        \State page $\gets$ req.page
        \State matchesSRI $\gets$ true
        \State sriDigest $\gets$ GetExpectedSRIDigest(page, req)
        \State matchesSRI $\gets$ sriDigest != null AND MatchesSRI(req, sriDigest)
        \State \Return not matchesSRI
        \EndProcedure
    \end{algorithmic}
\end{algorithm}

\begin{algorithm}[h]
    \small
    \caption{
        A sample integrity policy that denies requests if the remote server is in an unexpected location.
    }
    \label{lst:policy_location}
    \begin{algorithmic}[1]
        \State \textbf{Class} PolicyInfrastructureAttributes
        \State \ \ \ \ \textbf{Variable} locationSource = "example.com/.*"
        \State \ \ \ \ \textbf{Variable} locationTarget = ".*"
        \State \ \ \ \ \textbf{Variable} logic = IsInUnexpectedLocation
        \State \ \ \ \ \textbf{Variable} output = False
        \\

        \Procedure{IsInUnexpectedLocation}{req}
        \State distanceThreshold $\gets$ GetDistanceThreshold()
        \State ip $\gets$ GetRemoteIP(req)
        \State loc $\gets$ GetServerLocation(req)
        \If{IsExpectedCountry(loc.country)}
            \State \Return True
        \EndIf
        \Comment{Reference location can be adjusted}
        \State \Return GetDistance(loc) $>=$ distanceThreshold 
        \EndProcedure
    \end{algorithmic}
\end{algorithm}

\begin{algorithm}[h]
    \small
    \caption{
        A sample integrity policy that denies requests if modifications to client-side, or server-side, core files are detected.
    }
    \label{lst:policy_core_files}
    \begin{algorithmic}[1]
        \State \textbf{Class} PolicyCoreFiles
        \State \ \ \ \ \textbf{Variable} locationSource = "example.com/.*"
        \State \ \ \ \ \textbf{Variable} locationTarget = "example.com/core/.*"
        \State \ \ \ \ \textbf{Variable} logic = HasCoreFileChanged
        \State \ \ \ \ \textbf{Variable} output = False
        \\

        \Procedure{HasCoreFileChanged}{req}
        \State expected $\gets$ GetExpectedContent(req)
        \State actual $\gets$ GetActualContent(req)
        \State \Return expected != actual
        \EndProcedure
    \end{algorithmic}
\end{algorithm}

\section{Domain Sample Categorization}
\label{app:domain_sample_categorization}

The domain categories in \Cref{tab:domain_categories} were labeled by the free version of Google Gemini with the 2.5 Flash model.
Categories were assigned based on inferences made from domains mentioned in its training data and domain keyword analysis.

\begin{table}[t]
    \centering
    \caption{Domain sample categorization.}
    \resizebox{0.7\columnwidth}{!}{
        \begin{tabular}{l|r}
            \hline
            \textbf{Category} & \textbf{Count} \\
            \hline
            Technology/Software/Cloud & 127 \\
            Other/Uncategorized & 115 \\
            E-commerce/Retail & 50 \\
            News/Media/Publishing & 48 \\
            Government/Non-profit/Educational & 45 \\
            Entertainment/Adult Content & 56 \\
            Advertising/Marketing/Tracking & 39 \\
            Financial/Banking & 23 \\
            ISP/Telecommunications & 20 \\
            Social Media/Communication & 19 \\
            Gambling/Betting & 16 \\
            Health/Medical & 13 \\
            Forums/Communities & 12 \\
            Search/Portal & 10 \\
            Travel/Hospitality & 7 \\
            \hline
        \end{tabular}
    }
    \label{tab:domain_categories}
\end{table}

\section{Simulated Deployment}
\label{app:simulated_deployment}

\begin{figure}[t]
    \centering
    \includegraphics[width=1.0\columnwidth]{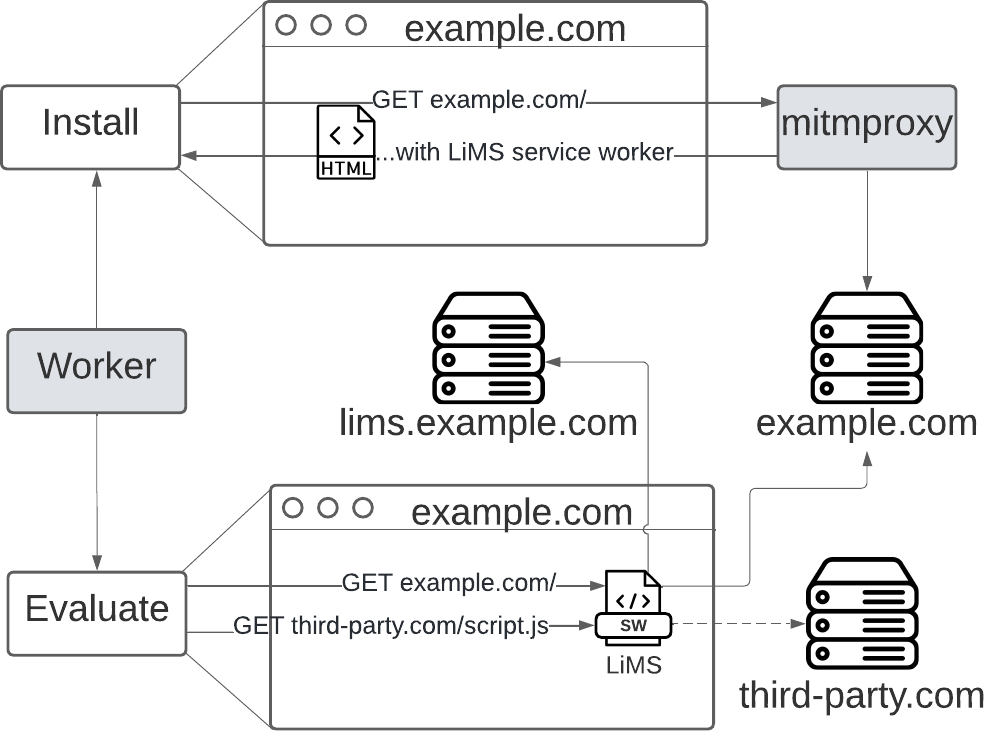}
    \caption{
        Diagram depicting the simulated deployment setup to bootstrap the \aliasSystemName{} client service worker for a live website.
    }
    \label{fig:diagram_simulated_deploy}
\end{figure}

We designed an evaluation pipeline to measure the performance overhead introduced by our system for an arbitrary website by simulating a deployment of \aliasSystemName{}: the pipeline deploys the backend system components and leverages a man-in-the-middle proxy to inject the client SW in the HTML responses from the target site, as depicted in \Cref{fig:diagram_simulated_deploy}.

For a given site, an evaluation worker visits it with Google Chrome for Testing \paramChromeVersion{} over a series of four different stages: the first with no service worker, and the next three corresponding to the \aliasSystemName{} prototype operating under each evaluation stage in \Cref{tab:eval_modes}, with at least 12 hours between two evaluation stages for any individual domain.
Each stage consists of visiting a domain, twice, with the configured evaluation mode.
The first visit bootstraps the client SW installation by leveraging a man-in-the-middle proxy to inject the client SW into HTML responses.
The browser's network cache and the service worker's local cache are cleared before relaunching the browser, without the proxy.
In this manner, the second visit is able to use the service worker that was previously injected and installed in the first visit.
In addition, we ensure that the browser components responsible for handling service workers are in a warm state by first visiting a website (not part of the evaluation set) that already uses a service worker, before visiting the target site.
After this, the browser proceeds to visit the target domain after a short delay, and reloads the page, while tracking these page load times to measure overhead.

\paragraph{Incremental Overhead}
The evaluation stages in \Cref{tab:eval_modes} progressively utilize more of the core functionality available in \aliasSystemName{}, and the last stage emulates a real deployment environment.
The first stage \fmtEvalStageZero{} measures the page load time with no service worker to function as a baseline.
The second stage \fmtEvalStageOne{} is designed to measure the overhead of a bare-bones service worker that does nothing but intercept network requests and immediately allow them to continue, and corresponds to reaching step 5 in \Cref{fig:diagram_system}.
The third stage \fmtEvalStageTwo{} is where we expect significant overhead --- for each new intercepted request, the client SW must send a \fmtQueryStatus{} request to the API server who immediately responds ``allowed'', corresponding to immediately after step 6 in \Cref{fig:diagram_system}.
The last stage \fmtEvalStageThree{} adds additional overhead by having the API perform database reads and determining whether requests should be allowed.

\paragraph{Default Policies}
We configure a set of default policies from \Cref{sec:03_blocks} for each target site, to simulate an actual deployment as closely as possible.
In particular, we install policies that verify that:
\begin{itemize}
    \item domains were not registered within the last 7 days,
    \item domains are not due to expire in the next 7 days,
    \item domains rank within the top 1M on Tranco, and
    \item the set of requests initiated by links do not change, and
    \item there are no TLS connection issues. 
\end{itemize}
We warm the \aliasSystemName{} system state by bootstrapping these policies, and their verifications, to ensure that there will always be valid verifications for links.
In addition, to avoid influencing the performance overhead evaluation, we specially modify the API server to always respond yes to client SW queries, regardless of the actual verifications.
During our performance evaluation experiments discussed in \Cref{subsec:04_perf}, we encountered several types of failed verification with these default policies.
The most common failures were caused by the default ranking policy: low-rank domains often contained links to other low-rank domains, with some not ranked in the top 1M on Tranco.
The other most common failure was caused by the default domain dropping policy which reported several third party domains that were due to expire.
The most popular among such domains, \code{adroll.com}, is likely an advertising service provider used by popular sites such as \code{x.com}, \code{macys.com}, \code{synology.com}, \code{datadoghq.com}, and \code{okta.com}, and it was re-registered in time.

\paragraph{Network Profiles}
Furthermore, as our system introduces additional network requests, we varied the network configurations to investigate whether there exists any undesired effects at slower speeds.
We performed our overhead measurements in \Cref{subsec:04_perf} at three common network speeds, while keeping the API server location constant, with an average latency of 5.8 ms and average packet loss of 0.53\% over 100 packets, as reported by \code{ping} from the evaluation nodes.
The first configuration is the unthrottled \numUnthrottledDownloadMedianMbps{} Mbps download and \numUnthrottledUploadMedianMbps{} Mbps upload speeds for the virtual machines in our institution's network; the second emulates the lower bounds of a Wi-Fi connection with 30 Mbps download and 15 Mbps upload speeds; the last emulates the lower bounds of a 5G low-band connection with 50 Mbps download and 10 Mbps upload speeds.
The unthrottled speeds vary depending on network activity from other tenants on the same physical host, and the numbers we present were obtained from computing the median of ten trials of the Ookla \code{speedtest-cli} tool~\cite{ooklaspeedtestcli} on each worker VM, on all VMs simultaneously.
We sourced the speeds for the Wi-Fi connection from its corresponding browser throttling profiles~\cite{mozillathrottling}, and the 5G low-band connection from industry reports~\cite{opensignal20225gexpreport} and prior 5G performance measurements~\cite{narayanan2021variegated}.

We used \code{wondershaper}~\cite{wondershaper}, a wrapper for the traditional Linux traffic shaper tool \code{tc}, to control the upload and download speeds from each of our evaluation nodes.
We opted to use \code{wondershaper} instead of the network throttling feature built into the browser because its throttling implementation does not automatically affect installed service workers~\cite{cdpnetworksw}, and we were unable to directly throttle the service worker context with the Chrome DevTools Protocol.

\end{document}